\documentclass[
 reprint,
 amssymb, amsmath,
 aip,cha
]{revtex4-1}

\usepackage[colorlinks=true,linkcolor=blue]{hyperref}
\usepackage{xspace}

\usepackage{graphicx}
\usepackage{subfig}
\usepackage{gensymb}
\usepackage{amsmath}
\usepackage{mathtools}

\usepackage[utf8]{inputenc}

\newcommand{\ket}[1]{{\ensuremath{|#1\rangle}\xspace}}
\newcommand{\bra}[1]{{\ensuremath{\langle #1|}\xspace}}
\newcommand{\elemm}[3]{{\ensuremath{\bra{#1}{#2}\ket{#3}}\xspace}}

\begin{document}

\title{
A Jeziorski-Monkhorst fully uncontracted Multi-Reference perturbative treatment I: 
principles, second-order versions and tests on ground state potential energy curves} 
\date{\today}

\author{Emmanuel Giner}
\email[Corresponding author. E-mail:]{E.Giner@fkf.mpg.de}
\affiliation{Max Planck Institue for Solid State Research, \\
Heisenbergstra{\ss}e 1, 70569, Germany}
\author{Celestino Angeli}
\affiliation{Dipartimento di Scienze Chimiche e Famaceutiche, \\Universita di Ferrara,
Via Fossato di Mortara 17,\\ I-44121 Ferrara, Italy}
\author{Yann Garniron}
\affiliation{Laboratoire de Chimie et Physique Quantiques,
(UMR 5626 of CNRS), IRSAMC, 
Universit\'e Paul Sabatier, 118 route de Narbonne,
F-31062 Toulouse Cedex, France}
\author{Anthony Scemama}
\affiliation{Laboratoire de Chimie et Physique Quantiques,
(UMR 5626 of CNRS), IRSAMC, 
Universit\'e Paul Sabatier, 118 route de Narbonne,
F-31062 Toulouse Cedex, France}
\author{Jean-Paul Malrieu}
\affiliation{Laboratoire de Chimie et Physique Quantiques,
(UMR 5626 of CNRS), IRSAMC, 
Universit\'e Paul Sabatier, 118 route de Narbonne,
F-31062 Toulouse Cedex, France}

\begin{abstract}
The present paper introduces a new multi-reference perturbation approach developed at second order, 
based on a Jeziorsky-Mokhorst expansion using individual Slater determinants as perturbers. 
Thanks to this choice of perturbers, an effective Hamiltonian may be built, 
allowing for the dressing of the Hamiltonian matrix within the reference space, assumed here to be a CAS-CI. 
Such a formulation accounts then for the coupling between the static and dynamic correlation effects.  
With our new definition of zeroth-order energies, these two approaches are strictly size-extensive provided that local orbitals are used, 
as numerically illustrated here and formally demonstrated in the appendix.
Also, the present formalism allows for the factorization of all double excitation operators, 
just as in internally contracted approaches, strongly reducing the computational cost 
of these two approaches with respect to other determinant-based perturbation theories. 
The accuracy of these methods has been investigated on ground-state potential curves up to full dissociation limits for 
a set of six molecules involving single, double and triple bond breaking. 
The spectroscopic constants obtained with the present methods are found to be in very good agreement 
with the full configuration interaction (FCI) results. 
As the present formalism does not use any parameter or numerically unstable operation, 
the curves obtained with the two methods are smooth all along the dissociation path. 
\end{abstract}
\maketitle

\section{Introduction\label{sec:intro}}
The research of the ground-state wave function of closed-shell molecules follows well-established paths. 
The perturbative expansions from the mean-field Hartree-Fock single determinant usually converge and may be used as basic tools, 
especially when adopting a mono-electronic zero-order Hamiltonian known as the M{\o}ller-Plesset Hamiltonian.\cite{mp} 
In this approach the wave function and the energy may be understood in terms of diagrams, 
which lead to the fundamental linked-cluster theorem.\cite{linked_goldstone} The understanding of the size-consistency problem led to the suggestion 
of the Coupled Cluster approximation,\cite{coester_1,coester_2,cizek,bartlett_1,barltett_2} 
which is now considered as the standard and most efficient tool in the study of such systems 
in their ground state, especially in its CCSD(T) version where linked corrections by triple excitations are added perturbatively.\cite{barltett_3} 
The situation is less evident when considering excited states, chemical reactions and molecular dissociations, since it then becomes impossible 
to find a relevant single determinant zero-order wave function. These situations exhibit an intrinsic Multi-Reference (MR) character. 
A generalized linked-cluster theorem has been established by Brandow,\cite{brandow} which gives a basis to the understanding of the size-consistency problem 
in this context, but the conditions for establishing this theorem are severe. They require a Complete Active Space (CAS) model space and 
a mono-electronic zero-order Hamiltonian. Consequently, the corresponding Quasi-Degenerate Perturbation Theory (QDPT) expansion cannot converge 
in most of the molecular MR situations\cite{intruder_States_1,intruder_States_2,intruder_States_3}. The research of theoretically satisfying (size-consistent) and numerically efficient MR treatments 
remains a very active field in Quantum Chemistry, as summarized in recent review articles concerning either perturbative\cite{mkpt2_1} or Coupled-Cluster\cite{bartlett_4} methods. 

The present work concentrates on the search of a new MR perturbative approach at second order (MRPT2).
Of course, pragmatic proposals have been rapidly formulated, consisting first in the identification of a reference model space, 
defined on the set of single determinants having large components in the desired eigenstates of the problem. 
Diagonalizing the Hamiltonian in this reference space delivers a zero-order wave function. Then one must define the vectors of the outer space 
to be used in the development and, in a perturbative context, choose a zero-order Hamiltonian. The simplest approach consists in using single 
determinants as outer-space eigenvectors, and this has been used in the CIPSI method\cite{cipsi_1,cipsi_2} which is iterative, 
increasing the model space from the selection of the perturbing determinant of largest coefficients and their addition to the model space. 
From a practical point of view this method is very efficient and is now employed to reach near exact Full Configuration Interaction (FCI) 
energies on small molecules,\cite{cucl2_1,cucl2_2} and also as trial wave function in the context of quantum Monte Carlo.\cite{3d_energies,f2_qmc,h2o_qmc,cipsi_opt} 
But the method suffers two main defects: i) 
it is not size-consistent and ii) it does not revise the model-space component of the wave function under the effect of its 
interaction with the outer-space. This last defect is avoided if one expresses the effect of the perturbation as a change 
of the matrix elements of the model space CI matrix, according to the Intermediate Effective Hamiltonian (IEH) theory,\cite{hint} 
as done in the state-specific\cite{hirao} or multi-state\cite{kirtman} versions.
Other methods which start from a CAS model space and use multi-determinantal outer-space vectors have been proposed later on and are now broadly used. 
The first one is the CASPT2 method,\cite{roos_1,roos_2} which employs a mono-electronic zero-order Hamiltonian. 
The method suffers from intruder state problems, to be cured in a pragmatic manner through the introduction of some parameters, 
and is not strictly size-consistent. The NEVPT2 method\cite{nevpt2_1,nevpt2_2,nevpt2_3} also uses multi-determinantal perturbers 
(defined in two different ways in its partially and strongly contracted versions), it makes use of a more sophisticated bi-electronic Hamiltonian 
(the Dyall Hamiltonian\cite{dyall}) to define the zero-order energies of these perturbers, it is parameter-free, intruder-state free, and size-consistent. 
Both methods are implemented in several popular codes, and use a contracted description of the model space component of the 
desired eigenfunction (fixed by the diagonalization of the Hamiltonian in the model space). Multi-State versions exist to give some flexibility 
to the model space component, in particular around weakly avoided crossings, but this flexibility is very limited\cite{mscaspt2,qdnevpt2}.
If one returns to methods using single-determinant perturbers, the origin of their size-inconsistency problem has been identified 
as due to the unbalance between the multi-determinant character of the zero-order wave function and the single determinant character of the perturbers.\cite{hmz} 
It is in principle possible to find size consistent formulations but they require rather complex formulations,\cite{mkmrpt2_1,mkmrpt2_2,mkmrpt2_3,ugamrpt2} 
and face some risk of numerical instabilities since they involve divisions by possibly small coefficients, the amplitudes of which may be small. 
Finally, one should mention a very recent approach based on the rewriting of the multi-reference linear coupled cluster equations in 
a stochastic framework of Full-CI Quantum Monte Carlo which also uses single Slater determinants as perturbers\cite{lmmrcc_fciqmc}.

The present paper is composed as follows. 
In Section \ref{theory}, the here-proposed formalism is presented, whose main features are:
\begin{enumerate}
 \item it considers a CAS model space (to achieve the strict separability requirement), usually obtained from a preliminary CASSCF calculation; 
 \item the perturbers are single determinants (the method is externally non-contracted, according to the usual terminology); 
 \item it is state-specific, and strictly separable when localized active MOs are used (see formal demonstration in the appendix); 
 \item it makes use of the Dyall Hamiltonian to define the excitation energies appearing in the energy denominators;
 \item it is based on a Jeziorski-Monkhorst\cite{jeziorsky_mokhorst} (JM) expression of the wave operator and proceeds through reference-specific partitionings 
       of the zero-order Hamiltonian, as it has been previously suggested in the so-called Multi-Partitionning\cite{mupa_1,mupa_2,mupa_3} (MUPA) 
       and also in the UGA-SSMPRT2.\cite{ugamrpt2} Consequently, it does not define a unique zero-order energy to the outer-space determinants;
 \item it can be expressed either as a second-order energy correction or as a dressing of the CAS-CI matrix, which offers a full flexibility in the treatment of the feed-back effect of the post-CAS-CI correlation on the model space component of the wave function; 
 \item the contributions of the various classes of excitations are easily identified (as in the CASPT2 and NEVPT2 methods);
 \item thanks to our definition of the zeroth-order energies, all processes involving double excitations can be treated 
       by using only the one- and two-boy density matrices, avoiding to loop on the perturbers; 
 \item given a set of molecular orbitals, it is parameter free and does not contain any threshold to avoid numerical instabilities.
\end{enumerate}
After having presented the working equations of the present formalism, section \ref{comparison} proposes a comparison with other existing 
MR approaches, such as some special cases of multi-reference coupled cluster (MRCC) and MRPT2. 
Then, Section \ref{results} presents the numerical results for the ground state potential energy curves of six molecules involving single, double and triple 
bond breaking with both the JM-MRPT2 and JM-HeffPT2 methods. A numerical test of size-extensivity is provided, 
together with the investigation of the dependency of the results on the locality of the active orbitals. 
Finally, Section \ref{concl} summarizes the main results and presents its tentative developments. 
The reader can find in the Section \ref{annex} a mathematical proof of strong separability of the JM-MRPT2 method. 

\section{Working equations for the perturbation and effective Hamiltonian at second order}
\label{theory}

\begin{figure}[h]
\includegraphics[width=0.8\columnwidth]{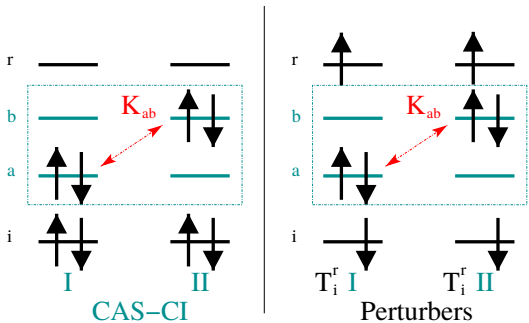}
\caption{Example of interactions: the two determinants of the CAS interact through a bi-electronic operator involving the two active orbitals $a$ and $b$, 
just as the two perturbers determinants generated by the same excitation operator $T_i^r$ on the two CAS determinants. }
\label{cas}
\end{figure}

As demonstrated previously by one of the present authors and his collaborators,\cite{hmz} 
the size-consistency problem in any multi-reference perturbative expansion using single Slater determinants as perturbers 
comes from the unbalanced zeroth order energies that occur in the denominators. 
More precisely, if the zeroth order wave function is a CAS-CI eigenvector, its energy is stabilized by all 
the interactions within the active space, whereas a perturber treated as a single Slater determinant 
does not benefit from these extra-diagonal Hamiltonian matrix elements. 
However, if one considers the set of perturber determinants created by the 
application of a given excitation operator on all Slater determinants of the CAS-CI wave function, 
most of the interactions found within the active space will occur within this set of perturbers 
(see Figure \ref{cas} for a pictorial example). Therefore, the use of linear combinations of Slater determinants as perturbers 
together with a bi-electronic zeroth order operator, as it is the case in the NEVPT2 framework which uses the Dyall 
zeroth order operator, leads to balanced energy differences and removes the size-consistency problem. 

On the basis of such considerations, the present work proposes an approach that uses single Slater determinants as 
perturbers and takes benefits of a new definition of energy denominators as expectation values 
of the Dyall zeroth order Hamiltonian over a specific class of linear combinations of Slater determinants. We first expose the definition of this perturbation theory, namely the JM-MRPT2, which
is strictly separable provided that local orbitals are used.

A large benefit from this new definition is that one may go beyond the sole calculation of the energy
and improve the reference wave function by taking into account, in a strictly size-consistent way, 
the correlation effects brought by the perturbers on the reference space. 
In a second step, we reformulate the approach as
a dressing of the Hamiltonian matrix within the set of Slater determinants
belonging to the reference wave function, which is diagonalized. This approach will be referred
to as the JM-HeffPT2 method.

\subsection{The JM-MRPT2 method}

\subsubsection{First-order perturbed wave function and second-order energy}

The formalism presented here is state specific and is not therefore restricted to ground state calculations.  
Nevertheless,  for the sake of clarity and compactness, we omit the index referring explicitly to a specific eigenstate.  

The zeroth order wave function $\ket{\psi^{(0)}}$ is assumed to be a CAS-CI eigenvector expanded on the set of 
reference determinants $\ket{\rm I}$: 
\begin{equation}
 \ket{\psi^{(0)}} = \sum_{\rm I\, \in \, CAS-CI} c_{\rm I } \ket{\rm I} 
\end{equation}
Such a wave function has a variational energy $e^{(0)}$: 
\begin{equation}
 e^{(0)} = \frac{\bra{\psi^{(0)}} H \ket{\psi^{(0)}} }{\bra{\psi^{(0)}} \psi^{(0)} \rangle }
\end{equation}
Starting from a normalized $\ket{\psi^{(0)}}$ (i.e. $\bra{\psi^{(0)}} \psi^{(0)} \rangle = 1$), 
we assume that the exact wave function can be expressed as: 
\begin{equation}
 \ket{\Psi} = \ket{\psi^{(0)}} + \sum_{\rm \mu \, \notin \, CAS-CI } c_{\mu} \ket{\mu}
\end{equation}
where here the $\ket{\mu}$ are all possible Slater determinants not belonging to the CAS-CI space. 
One should notice that such form is in principle not exact, as some changes of the coefficients 
within the CAS-CI space can formally occur when passing from the CAS-CI eigenvector to the FCI one, 
but such an approximated  form for the exact wave function is the basis of many MRPT2 approaches 
like NEVPT2, CASPT2 or CIPSI. 

As in any projection technique, the exact energy can be obtained by projecting the Schr\"odinger 
equation on $\ket{\psi^{(0)}}$: 
\begin{equation}
 \begin{aligned}
 E & = \bra{\psi^{(0)}} H \ket{\Psi} \\
%  & = \bra{\psi^{(0)}} H \ket{\psi^{(0)}} 
   & = e^{(0)} + \sum_{\rm \mu \, \notin \, CAS-CI} c_{\mu} \, \bra{\psi^{(0)}} H \ket{\mu}
 \end{aligned}
\end{equation}
and one only needs to compute the coefficients of the $\ket{\mu}$ that interact with $\ket{\psi^{(0)}}$, 
which consist in all individual Slater determinants being singly or doubly excited with respect to any Slater determinant 
in the CAS-CI space. From now on we implicitly refer to $\ket{\mu}$ as any single Slater determinant belonging to 
such a space. 

The coefficients $c_{\mu}$ are then written according to the JM ansatz\cite{jeziorsky_mokhorst}, 
whose general expression for wave function is not explicitly needed here, and will  be therefore given in the section \ref{linkmrcc} 
when the comparison of  the present method with other multi-reference methodologies will be investigated.  
The JM ansatz introduces the genealogy of the coefficients $c_{\mu}$ with respect to the Slater determinants within the CAS-CI space:
\begin{equation}
 \label{jmdef1}
 c_{\mu} = \sum_{\rm I } c_{\rm I} \,\, t_{\rm I \mu}
\end{equation}
where the quantity $t_{\rm I \mu}$ is the excitation amplitude related to the excitation process $T_{\rm I \mu}$ that leads from $\ket{\rm I}$ 
to $\ket{\mu }$: 
\begin{equation}
 T_{\rm I \mu} \ket{\rm I } = \ket{\mu}
\end{equation}
Here, we restrict $T_{\rm I \mu} $ to be a single or double excitation operator.  
Within this JM formulation of $c_{\mu}$, a very general first order approximation of the amplitudes $t_{\rm I \mu}^{(1)}$ can be expressed as:
\begin{equation}
 \label{jmdef2}
 t_{\rm I \mu}^{(1)} = \frac{\bra{\rm I} H \ket{\mu}}{\Delta E^{(0)}_{\rm I \mu}}
\end{equation}
where the excitation energy  $\Delta E^{(0)}_{\rm I \mu}$ depends explicitly of the couple $\left(\ket{\rm I},\,\ket{\mu}\right)$. 
Such a definition is different from other determinant-based MRPT2 like the CIPSI or shifted-$B_k$ where the excitation energy does 
not depend on the couple $\left(\ket{\rm I},\,\ket{\mu}\right)$ but only on the $\ket{\mu}$.  
With this definition of $t_{\rm I \mu}^{(1)}$, one can write the second-order correction to the energy $e^{(2)}$ as:
\begin{equation}
 \begin{aligned}
 \label{e2pert}
 e^{(2)} & = \elemm{\psi^{(0)}}{H}{\psi^{(1)}} \\ 
   & =  \sum_{\mu } \sum_{\rm I } c_{\rm I} \frac{\elemm{\rm I}{H}{\mu} }{\Delta E^{(0)}_{{\rm I \mu}}} \elemm{\psi^{(0)}}{H}{\mu} \\
                  & = \sum_{\mu} \sum_{\rm I\, J} c_{\rm I} \,\,\frac{\bra{\rm I} H \ket{\mu}\elemm{\mu }{H}{\rm J}}{\Delta E^{(0)}_{\rm I \mu}}\,\, c_{\rm J} 
 \end{aligned}
\end{equation}
and the total second-order energy $E^{(2)}$:
\begin{equation}
 \label{defE2}
 \begin{aligned}
 E^{(2)} & = \elemm{\psi^{(0)}}{H}{\psi^{(0)}} + \elemm{\psi^{(0)}}{H}{\psi^{(1)}} \\
         & = e^{(0)} + e^{(2)}
 \end{aligned}
\end{equation}

\subsubsection{Definition of the energy denominators}

The first-order wave function can be written explicitly in terms of the excitation operators $T_{\rm I \mu}$:
\begin{equation}
 \begin{aligned}
 \ket{\psi^{(1)}} & = \sum_{\mu} c_{\mu}^{(1)} \ket{\mu} \\
                  & = \sum_{\mu} \sum_{\rm I} c_{\rm I} \,\,\frac{\bra{\rm I} H \,\,T_{\rm I \mu}\ket{\rm I}}{\Delta E^{(0)}_{\rm I \mu}}\,\, T_{\rm I \mu} \ket{\rm I }
 \end{aligned}
\end{equation}
However, one can notice that 
\begin{enumerate}
 \item the excitation operators $T_{\rm I \mu}$ do not explicitly depend on $\ket{\rm I}$ as they are general single or double 
       excitation operators, just as in the Hamiltonian for instance; 
 \item a given excitation operator $T$ contributes to the coefficients of several $\ket{\mu}$
  ($T_{\rm I \mu}$ = $T_{\rm J \nu}$ = $T$); 
 \item the application of all the single and double excitation operators $T$ on each $\ket{\rm I}$ generates the entire set 
       of $\ket{\mu}$ as the reference is a CAS. 
\end{enumerate}
Therefore one can rewrite the first-order perturbed wave function directly in term of excitation operators $T$ applied 
on the each CAS-CI Slater determinant as:
\begin{equation}
 \ket{\psi^{(1)}} = \sum_{T} \ket{\psi^{(1)}_T} 
\end{equation}
where the $\ket{\psi^{(1)}_T}$ is the part of the first-order wave function associated with the excitation process $T$: 
\begin{equation}
 \label{def_psit}
 \ket{\psi^{(1)}_T} = \sum_{\rm I} c_{\rm I} \,\,\frac{\bra{\rm I} H \,\,T\ket{\rm I}}{\Delta E^{(0)}_{{\rm I} \,\,T {\rm I}}} \,\,T \ket{\rm I }
\end{equation}

In order to fully define our perturbation theory and intermediate Hamiltonian theory, 
one needs to select an expression for the energy denominators occurring in the definition of $\ket{\psi^{(1)}_T}$. 
We propose to take a quantity that does not depend explicitly on the reference determinant $\ket{\rm I}$ 
but only depends on the excitation process $T$: 
\begin{equation}
 \label{def_deltae_uniq}
  \Delta E^{(0)}_{{\rm I} \,\,T {\rm I}}  = \Delta E^{(0)}_T \quad \forall \,\, {\rm I}
\end{equation}
Consequently, in the expression of $\ket{\psi^{(1)}_T}$ (see Eq.~\eqref{def_psit}), the energy denominator can be factorized:
\begin{equation}
 \label{def_deltae_uniq}
  \begin{aligned}
   \ket{\psi^{(1)}_T} & = \frac{1}{\Delta E^{(0)}_T} \sum_{\rm I} c_{\rm I} \,\,\bra{\rm I} H \,\,T\ket{\rm I} \,\,T \ket{\rm I } \\
                      & = \frac{1}{\Delta E^{(0)}_T} \ket{\tilde{\psi}^{(1)}_T}
  \end{aligned}
\end{equation}
where $\ket{\tilde{\psi}^{(1)}_T}$ is simply:
\begin{equation}
 \ket{\tilde{\psi}^{(1)}_T} = \sum_{\rm I} c_{\rm I} \,\,\bra{\rm I} H \,\,T\ket{\rm I} \,\,T \ket{\rm I }
\end{equation}
Also, one can notice that, as $\ket{\tilde{\psi}^{(1)}_T}$ and $\ket{\psi^{(1)}_T}$ differ by a simple constant factor, they have the same normalized 
expectation values:
\begin{equation}
   \frac{\bra{{\psi}^{(1)}_T} H^{D} \ket{{\psi}^{(1)}_T}}{\bra{{\psi}^{(1)}_T} {\psi}^{(1)}_T\rangle} = 
   \frac{\bra{\tilde{\psi}^{(1)}_T} H^{D} \ket{\tilde{\psi}^{(1)}_T}}{\bra{\tilde{\psi}^{(1)}_T} \tilde{\psi}^{(1)}_T\rangle}
\end{equation}
Then, the excitation energy $\Delta E^{(0)}_T $ is simply taken as the difference of the 
normalized expectation values of the Dyall Hamiltonian $H^{D}$ over $\ket{\psi^{(0)}}$ and $\ket{{\tilde{\psi}}^{(1)}_T}$: 
\begin{equation}
 \label{def_deltae}
  \begin{aligned}
  \Delta E^{(0)}_{T }  & = \frac{\bra{\psi^{(0)}} H^{D} \ket{\psi^{(0)}}}{\bra{\psi^{(0)}} \psi^{(0)}\rangle}  -  \frac{\bra{\tilde{\psi}^{(1)}_T} H^{D} \ket{\tilde{\psi}^{(1)}_T}}{\bra{\tilde{\psi}^{(1)}_T} \tilde{\psi}^{(1)}_T\rangle} \\
                       & =  \frac{\bra{\psi^{(0)}} H^{D} \ket{\psi^{(0)}}}{\bra{\psi^{(0)}} \psi^{(0)}\rangle}  -  \frac{\bra{{\psi}^{(1)}_T} H^{D} \ket{{\psi}^{(1)}_T}}{\bra{{\psi}^{(1)}_T} {\psi}^{(1)}_T\rangle} 
  \end{aligned}
\end{equation}
This ensures the strong separability when localized orbitals are used, as will be illustrated numerically in the section \ref{results}.  

The Dyall Hamiltonian is nothing but the exact Hamiltonian over the active orbitals, and a M{\o}ller-Plesset type operator 
over the doubly occupied and virtual orbitals. If one labels $a,\, b,\, c,\, d$ the active spin-orbitals, $i,\,j$ the 
spin-orbitals that are always occupied and $v,\,r$ the virtual spin-orbitals, 
the Dyall Hamiltonian can be written explicitly as:  
\begin{equation}
 H^D = H^D_{iv} + H^D_a \\
\end{equation}
\begin{equation}
% \left\{
  \begin{dcases}
 H^D_a = \sum_{ab} h^{\rm eff}_{ab} a^{\dagger}_a a_{b} + \frac{1}{2}\sum_{abcd}  \, (ad|bc) \,\,a^{\dagger}_a a^{\dagger}_b a_c a_d \\
 \label{dyall_inact}
  H^D_{iv} =  \sum_{i} \, \epsilon_i \,\,a^{\dagger}_i a_i + \sum_{v} \, \epsilon_v \,\, a^{\dagger}_v a_{v} + C 
  \end{dcases}
%  \right.
\end{equation}
where the $\epsilon_i$ and $\epsilon_v$ are defined as the spin-orbital energies associated with the density given 
by $\ket{\psi^{(0)}}$, and the effective active one-electron operator 
$h^{\rm eff}_{ab} = \bra{a} h + \sum_{i} \left(J_i - K_i\right) \ket{b}$. With a proper choice of the constant $C$ 
in Eq.~\eqref{dyall_inact},
\begin{equation}
C = \sum_i \bra{i} h \ket{i} + \frac{1}{2} \sum_{i,j} \left((ii|jj)-(ij|ij) \right),
\end{equation}
one has: 
\begin{equation}
 \frac{\bra{\psi^{(0)}} H^{D} \ket{\psi^{(0)}}}{\bra{\psi^{(0)}} \psi^{(0)}\rangle} = \frac{\bra{\psi^{(0)}} H \ket{\psi^{(0)}}}{\bra{\psi^{(0)}} \psi^{(0)}\rangle} = e^{(0)}
\end{equation}
Because the Dyall Hamiltonian acts differently on the active and inactive-virtual orbitals,  
the excitation energy $\Delta E^{(0)}_T$ is the sum of an excitation energy $\Delta E^{(0)\, iv}_{T}$ 
associated with the inactive and virtual orbitals and of an excitation energy $\Delta E^{(0)\,a}_{T}$  
associated with the active orbitals: 
\begin{equation}
 \Delta E^{(0)}_T = \Delta E^{(0)\,a}_{T} + \Delta E^{(0)\, iv}_{T}
\end{equation}
Also, it is useful to differentiate the active part from the inactive-virtual part of the excitation $T$: 
\begin{equation}
 T = T_a T_{iv}
\end{equation}
The inactive-virtual excitation energy $\Delta E^{(0)\, iv}_{T}$ is simply: 
\begin{equation}
 \Delta E^{(0)\, iv}_{T} = \sum_{i \, \in \, T } \epsilon_i - \sum_{v\, \in \, T} \epsilon_v
\end{equation}
where $i \, \in \, T$ and $v\, \in \, T$ refer to, respectively, the inactive and virtual spin-orbitals 
involved in the excitation operator $T$. 
Conversely, the active excitation energy $\Delta E^{(0)\, a}_{T}$ has a more complex expression, namely:
\begin{widetext}
\begin{equation}
 \label{deltae}
 \Delta E^{(0)\,a}_{T} = e^{(0)} -
\frac{\sum_{\rm I\,\, J} \left(c_{\rm I}  \bra{\rm I}H \, T \ket{\rm I}\right) \,\, \bra{\rm I}T^{\dagger}_a\,  H^D\, T_a \ket{\rm J} \,\, \left(c_{\rm J} \bra{\rm J}H \, T \ket{\rm J}\right)   }
{ \sum_{\rm I} \left(c_{\rm I}  \bra{\rm I}H \, T \ket{\rm I}\right)^2  \bra{\rm I}T^{\dagger}\,\, T \ket{\rm I} }.
\end{equation}
\end{widetext}

\subsubsection{Practical consequences: the difference between single and double excitation operators}
From Eq.~\eqref{deltae}, one must differentiate the class of the pure single excitation operators from 
the pure double excitation operators. 
For the sake of clarity, we define the spin-adapted bielectronic integrals $((mn|pq))$ as:
\begin{equation}
 ((mn|pq)) =
  \begin{cases}
   (mn|pq)            & \text{if } \sigma(m,p) \ne \sigma(n,q) \\
   (mn|pq) - (mp|nq)  & \text{if } \sigma(m,p) = \sigma(n,q) \\
  \end{cases}
\end{equation}
where $\sigma(m,p)$ is the spin variable of the spin orbitals $m$ and $p$. 
If one considers a given double excitation involving four different spin orbitals $m$, $n$, $p$ and $q$:
\begin{equation}
 T_{mp}^{nq} = a^{\dagger}_{n} a^{\dagger}_q a_p a_m \qquad m\ne n \ne p \ne q
\end{equation}
one can notice that the Hamiltonian matrix elements associated with this double excitation only depend, up to a phase factor, on the four indices
$m,n,p,q$ involved in the $T_{mp}^{nq}$. Indeed, if $T_{mp}^{nq}$ is possible on both $\ket{\rm I}$ and $\ket{\rm J}$, one has:
\begin{equation}
 \begin{aligned}
 \elemm{\rm I}{H\,\,T_{mp}^{nq} }{\rm I} & =  ((mn|pq)) \,\,   \elemm{\rm I}{ \left(T_{mp}^{nq}\right)^{\dagger} T_{mp}^{nq}}{\rm I} \\
 \elemm{\rm J}{H\,\,T_{mp}^{nq} }{\rm J} & =  ((mn|pq)) \,\,   \elemm{\rm J}{ \left(T_{mp}^{nq}\right)^{\dagger} T_{mp}^{nq}}{\rm J} 
 \end{aligned}
\end{equation} 
and as
\begin{equation}
 \begin{aligned}
  \elemm{\rm I}{ \left(T_{mp}^{nq}\right)^{\dagger} T_{mp}^{nq}}{\rm I}   & =  \elemm{\rm J}{ \left(T_{mp}^{nq}\right)^{\dagger} T_{mp}^{nq}}{\rm J} \\
  & =  1
 \end{aligned}
\end{equation} 
it becomes:
\begin{equation}
 \elemm{\rm J}{H\,\,T_{mp}^{nq} }{\rm J}  =  \elemm{\rm I}{H\,\,T_{mp}^{nq} }{\rm I}
\end{equation} 
Therefore, as the hamiltonian matrix elements of type $\elemm{\rm J}{H\,\,T_{mp}^{nq} }{\rm J}$ can be factorized both in the numerator and the dominator 
of the expression of the active part of the excitation energy (see Eq. \eqref{deltae}). 
Finally,  the expression of the active part of the excitation energy for a given double excitation $T_{mp}^{nq}$ is simply:
\begin{equation}
 \label{deltae_double}
 \begin{aligned}
 \Delta E^{(0)\, a}_{T_{mp}^{nq} } & = e^{(0)} - \frac{\sum_{\rm I\,\, J} c_{\rm I}   \,\, \bra{\rm I}T^{\dagger}_a\,  H^D\, T_a \ket{\rm J} \,\, c_{\rm J}  }{ \sum_{\rm I} c_{\rm I}^{2}  \bra{\rm I}T_a^{\dagger}\,\, T_a \ket{\rm I} }  \\
                       & = e^{(0)} - 
 \frac{\bra{\psi^{(0)}}T^{\dagger}_a\, H^D\, T_a\ket{\psi^{(0)}}}{\bra{\psi^{(0)}}T^{\dagger}_a T_a\ket{\psi^{(0)}}} 
 \end{aligned}
\end{equation}
As a consequence, the amplitudes $t_{{\rm I} T_{mn}^{qp} {\rm I}}$  and $t_{{\rm J} T_{mn}^{qp} {\rm J}}$ associated with the same excitation 
$T_{mn}^{qp}$ for different parents $\ket{\rm I}$ and $\ket{\rm J}$ are also equal:
\begin{equation}
 \begin{aligned}
 t_{{\rm I} T_{mn}^{qp} {\rm I}} & = \frac{\elemm{\rm I}{H\,\,T_{mn}^{qp} }{\rm I}}{\Delta E^{(0)}_{T_{mn}^{qp}}}\\
 t_{{\rm J} T_{mn}^{qp} {\rm J}} & = \frac{\elemm{\rm J}{H\,\,T_{mn}^{qp} }{\rm J}}{\Delta E^{(0)}_{T_{mn}^{qp}}}
 \end{aligned}
\end{equation}
and one can define a unique excitation operator ${\mathcal{T}_{mn}^{qp}}^{(1)}$ 
which does not depend on the reference determinant on which it acts. 
The explicit form of the reference-independent excitation operator ${\mathcal{T}_{mn}^{qp}}^{(1)}$ is
\begin{equation}
 \label{eq:tmnpq}
 {\mathcal{T}_{mn}^{qp}}^{(1)} = \frac{((mq|np))}{\Delta E^{(0)}_{T_{mn}^{qp}}} a^{\dagger}_{q} a^{\dagger}_p a_n a_m 
\end{equation}
%This expression involves only the active part of the Dyall Hamiltonian, and can be interpreted 
%as the energetic cost related to the excitation $T_a$. We will explicit the type of energy differences for each 
%class of excitations.\\  

In the case where $T$ is a pure single excitation operator, the term
$\bra{\rm I}H \, T \ket{\rm I}$ 
may strongly depend on $\ket{\rm I}$ and Eq.~\eqref{deltae} cannot be simplified.

\subsubsection{Precaution for spin symmetry}
As the formalism proposed here deals with Slater determinants, it cannot formally ensure to provide spin eigenfunctions. 
In order to ensure the invariance of the energy with the $S_z$ value of a given spin multiplicity, 
we introduced a slightly modified version of the Dyall Hamiltonian which does not consider: 
\begin{enumerate}
 \item any exchange terms in the Hamiltonian matrix elements when active orbitals are involved 
 \item any exchange terms involving two electrons of opposite spins (namely $a^{\dagger}_{b\alpha} a^{\dagger}_{a\beta} a_{b\beta} a_{a\alpha} $ and $a^{\dagger}_{b\beta} a^{\dagger}_{a\alpha} a_{b\alpha} a_{a\beta} $)
\end{enumerate}

\subsection{The JM-HeffPT2 method}

An advantage of a determinant-based multi-reference perturbation theory  
is that it can be easily written as a dressing of the Hamiltonian matrix within the reference space. 
Starting from the Schr\"odinger equation projected on a given reference determinant $\ket{\rm I}$ one has: 
\begin{equation}
 c_{\rm I} \elemm{\rm I}{H}{\rm I} + \sum_{\rm J \ne I} c_{\rm J} \elemm{\rm I}{H}{\rm J} 
 \sum_{\mu} c_{\mu}^{(1)} 
 \elemm{\rm I}{H}{\mu} = E^{(2)} c_{\rm I}.
\end{equation}
Using the expression for the first order coefficients $c_{\mu}^{(1)}$, it becomes:
\begin{multline}
 \label{eigv}
 c_{\rm I} \left( \elemm{\rm I}{H}{\rm I} + \sum_{\mu} \frac{\elemm{\rm I}{H}{\mu}^2 }{\Delta E^{(0)}_{{\rm I \mu}}} \right) + \\  \sum_{\rm J \ne I} c_{\rm J} \left( \elemm{\rm I}{H}{\rm J} 
  + \frac{\elemm{\rm I}{H}{\mu}\elemm{\mu}{H}{\rm J}}{\Delta E^{(0)}_{{\rm J \mu}}} \right)
   = E^{(2)} c_{\rm I}.
\end{multline}
Therefore, one can define an non-Hermitian operator $\Delta H^{(2)}$: 
\begin{equation}
 \elemm{\rm I}{\Delta H^{(2)}}{\rm J} = \sum_{\mu } \frac{\elemm{\rm I}{H}{\mu}\elemm{\mu}{H}{\rm J}}{\Delta E^{(0)}_{{\rm J \mu}}}
\end{equation}
and a dressed Hamiltonian $\mathcal{H}$ as:
\begin{equation}
 \elemm{\rm I}{\mathcal{H}^{(2)}}{\rm J} = \elemm{\rm I}{H}{\rm J} + \elemm{\rm I}{\Delta H^{(2)}}{\rm J} 
\end{equation}
such that Eq.~\eqref{eigv} becomes a non-symmetric linear eigenvalue equation within the CAS-CI space: 
\begin{equation}
 \label{dresseigv}
 c_{\rm I} \elemm{\rm I}{\mathcal{H}^{(2)}}{\rm I} + \sum_{\rm J \ne I} c_{\rm J} \elemm{\rm I}{\mathcal{H}^{(2)}}{\rm J}  = 
 E^{(2)} c_{\rm I}
\end{equation}
The second-order correction to the energy $e^{(2)}$ can be simply obtained as the expectation value of $\Delta H^{(2)}$ 
over the zeroth-order wave function: 
\begin{equation}
 \begin{aligned}
 e^{(2)} & = \elemm{\psi^{(0)}}{\Delta H^{(2)}}{\psi^{(0)}} \\
   & =  \sum_{\mu } \sum_{\rm I \, J} c_{\rm I} \,\,\frac{\elemm{\rm I}{H}{\mu} \elemm{\mu}{H}{\rm J} }{\Delta E^{(0)}_{{\rm J \mu}}} \,\, c_{\rm J} .
 \end{aligned}
\end{equation}
Finally, one can define a Hermitian operator $\tilde{H}^{(2)}$: 
\begin{equation}
 \elemm{\rm I}{\tilde{H}^{(2)}}{\rm J} = \frac{1}{2} \left(\elemm{\rm I}{\mathcal{H}^{(2)}}{\rm J} + \elemm{\rm J}{\mathcal{H}^{(2)}}{\rm I}\right)
\end{equation}
and a corresponding eigenpair ($\ket{\tilde{\Psi}_2}$, $\tilde{E}^{(2)}$) verifying:
\begin{equation}
 \label{defHeff2}
 \tilde{H}^{(2)} \ket{\tilde{\Psi}_{2}} = \tilde{E}^{(2)} \ket{\tilde{\Psi}_{2}}
\end{equation}
The diagonalization of such a Hamiltonian allows then to improve the CAS-CI wave function
by treating the coupling that can exist between the correlation effects within and outside 
the CAS-CI space.

\section{Links with other multi-reference methods}
\label{comparison}

\subsection{Strongly contracted NEVPT2}
It is interesting to understand the similarities and differences between the present JM-MPRT2 
and other strictly size-consistent MRPT2 methods,   like the NEVPT2 and specially its strongly contracted variant (SC-NEVPT2). 
The first important similarity is that both the JM-MRPT2 and the NEVPT2 methods use the Dyall Hamiltonian. 
Then, the JM-MRPT2 uses perturbers that are individual Slater determinants, whereas the NEVPT2 uses linear combinations of Slater determinants.
However, in the SC-NEVPT2, the contraction coefficients are closely related to the Hamiltonian matrix elements, just as in the JM-MRPT2 method. 
In order to better understand the differences between the SC-NEVPT2 and JM-MRPT2, let us take a practical example. 
Here, $i,j$ are inactive spin-orbitals, $a,b$ are active spin-orbitals and $r,s$ are virtual spin-orbitals. 
Considering a given semi-active double excitation $T_{ij}^{av} = a^{\dagger}_a a^{\dagger}_v a_j a_i$, the first-order amplitude ${t_{ij}^{av}}^{(1)}$ 
associated with $T_{ij}^{av}$ in the JM-MRPT2 formalism is given by:
\begin{equation}
 {t_{ij}^{av}}^{(1)} = \frac{((ia|jv))}{\epsilon_i + \epsilon_j - \epsilon_v + \Delta E^{(0)}_{a^{\dagger}_a}}
\end{equation}
where the active part of the excitation energy $\Delta E^{(0)}_{a^{\dagger}_a}$ directly comes from Eq.~\eqref{deltae_double}: 
\begin{equation}
 \label{deltaea}
 \Delta E^{(0)}_{a^{\dagger}_a} =  e^{(0)} - 
 \frac{\bra{\psi^{(0)}}a_a\, H^D\, a^{\dagger}_a\ket{\psi^{(0)}}}{\bra{\psi^{(0)}}a_a a^{\dagger}_a\ket{\psi^{(0)}}}.
\end{equation}
Note that such a quantity can be thought as an approximation of the electron affinity of the molecule, 
as it is the change in energy when one introduces ``brutally'' an electron in spin orbital $a$ without relaxing the wave function.  
Consequently, as it has been emphasized in section \ref{linkmrcc}, one can consider the part of the first-order perturbed wave function generated by the 
excitation $T_{ij}^{av}$: 
\begin{equation}
  \ket{\psi^{(1)}_{T_{ij}^{av}}} = \sum_{\rm I} c_{\rm I} \,\, {t_{ij}^{av}}^{(1)} \,\, T_{ij}^{av} \ket{\rm I}
\end{equation}
which turns out to be:
\begin{equation}
 \label{defpsit}
 \begin{aligned}
  \ket{\psi^{(1)}_{T_{ij}^{av}}} & = \frac{((ia|jv))}{\epsilon_i + \epsilon_j - \epsilon_v + \Delta E^{(0)}_{a^{\dagger}_a}} \sum_{\rm I} c_{\rm I} \,\, 
  T_{ij}^{av} \,\, \ket{\rm I} \\
                                 & = \frac{((ia|jv))}{\epsilon_i + \epsilon_j - \epsilon_v + \Delta E^{(0)}_{a^{\dagger}_a}} T_{ij}^{av} 
  \ket{\psi^{(0)}}
 \end{aligned}
\end{equation}
In the SC-NEVPT2 framework, one does not consider explicitly a given $T_{ij}^{av}$ but has to consider a unique excitation $\mathcal{T}_{ij}^{v}$ 
which is a linear combination of all possible $T_{ij}^{av}$ for all active spin orbitals $a$, with proper contraction coefficients. 
To be more precise, the first-order perturbed wave function associated with $\mathcal{T}_{ij}^{v}$ is:
\begin{equation}
 \ket{\psi^{(1)}_{\mathcal{T}_{ij}^{v}}} = \frac{1}{\Delta E^{(0)}_{\mathcal{T}_{ij}^{v}}} \sum_{a} \,((ia|jv)) \,\, T_{ij}^{av} \,\, \ket{\psi^{(0)}}
\end{equation}
where the excitation energy $\Delta E^{(0)}_{\mathcal{T}_{ij}^{v}}$ associated with $\mathcal{T}_{ij}^{v}$ 
is unique for all the excitation operators $T_{ij}^{av}$, and can be thought as an average excitation energy over all $a$. 
Consequently, one can express the part of $\ket{\psi^{(1)}_{\mathcal{T}_{ij}^{v}}}$ that comes from the $T_{ij}^{av}$ as:
\begin{equation}
 \label{defpsit_nevpt2}
 \ket{\psi^{(1)}_{T_{ij}^{av}}}^{\rm (SC-NEVPT2)} = \frac{((ia|jv))}{\Delta E^{(0)}_{\mathcal{T}_{ij}^{v}}}  T_{ij}^{av} \,\, \ket{\psi^{(0)}}
\end{equation}
which we can compare to Eq.~\eqref{defpsit} in the case of the JM-MRPT2 method. 
Then, the only difference between the SC-NEVPT2 and the JM-MRPT2 is the definition of the excitation energy occurring 
in Eqs.~\eqref{defpsit} and~\eqref{defpsit_nevpt2}. In the SC-NEVPT2 method, the excitation energy $\Delta E^{(0)}_{\mathcal{T}_{ij}^{v}}$ 
is closely related to the excitation energy  defined in JM-MRPT2:
\begin{equation}
 \begin{aligned}
 \Delta E^{(0)}_{\mathcal{T}_{ij}^{v}} & = e^{(0)} - 
 \frac{\bra{\psi^{(1)}_{\mathcal{T}_{ij}^{v}}}\, H^D\, \ket{\psi^{(1)}_{\mathcal{T}_{ij}^{v}}}}{\bra{\psi^{(1)}_{\mathcal{T}_{ij}^{v}}} \psi^{(1)}_{\mathcal{T}_{ij}^{v}}\rangle}  \\
                                       & = \epsilon_i + \epsilon_j - \epsilon_v + \Delta E^{(0)\rm SC-NEVPT2}_{a^{\dagger}}
 \end{aligned}
\end{equation}
where the quantity $\Delta E^{(0)\rm SC-NEVPT2}_{a^{\dagger}}$ is the same for all active orbitals and defined as:
\begin{widetext}
\begin{equation}
 \Delta E^{(0)\rm SC-NEVPT2}_{a^{\dagger}} =  e^{(0)} - 
 \frac{\sum_{a} \sum_{b}  ((ia|jv)) ((ib|jv)) \elemm{\psi^{(0)}}{ a_b  \,\,  H^D  \, a^{\dagger}_a}{\psi^{(0)}}}{\sum_a ((ia|jv))^2 \bra{\psi^{(0)}}a_a a^{\dagger}_a\ket{\psi^{(0)}}}.
\end{equation}
\end{widetext}
Under this perspective, one sees that the quantity $\Delta E^{(0)\rm SC-NEVPT2}_{a^{\dagger}}$ is related to $\Delta E^{(0)}_{a^{\dagger}_a}$ defined in Eq.~\eqref{deltaea}: 
 \begin{itemize}
 \item in the JM-MRPT2 method, the quantity $\Delta E^{(0)}_{a^{\dagger}_a}$ explicitly refers to the ``brutal'' addition 
       of en electron in orbital $a$, whatever the inactive orbitals $i,j$ or virtual orbitals $v$ 
       involved in $T_{ij}^{av}$; 
 \item the quantity $\Delta E^{(0)\rm SC-NEVPT2}_{a^{\dagger}}$ involved in the SC-NEVPT2 is an average electronic affinity 
       over all possible excitation processes $a^{\dagger}_a$ within the active space, but keeping a trace 
       of the inactive and virtual excitation processes involved in $T_{ij}^{av}$ thanks to the interaction $(ia|jv)$. 
\end{itemize}
Consequently, the quantity $\Delta E^{(0)\rm SC-NEVPT2}_{a^{\dagger}}$ contains also the interactions between the various
$a^{\dagger}_a\,\,\ket{\psi^{(0)}}$. 
To summarize, on one hand, the JM-MRPT2 gives a different but rather crude excitation energy for each $T_{ij}^{av}$, 
and on the other hand the SC-NEVPT2 has a unique and sophisticated excitation energy for all $T_{ij}^{av}$.  
Of course, one can extend this comparison to all the other classes of double excitations.

\subsection{Multi-reference coupled cluster methods}
\label{linkmrcc}
The present formalism has also several links with other multi-reference methods. 
First of all, as it uses a JM genealogical definition for the coefficients 
$c_{\mu}^{(1)}$ (see Eqs.~\eqref{jmdef1} and~\eqref{jmdef2}), the wave function corrected at first order $\ket{\Psi^{(1)}}$ can be written as: 
\begin{equation}
 \label{defPsi1}
 \begin{aligned}
 \ket{\Psi^{(1)}} & = \ket{\psi^{(0)}} + \ket{\psi^{(1)}} \\
                  & = \sum_{\rm I} c_{\rm I} \,\, \ket{\rm I} + \sum_{\mu} \sum_{\rm I} c_{\rm I} \,\, t_{\rm I \mu}^{(1)}\,\, T_{\rm I \mu} \ket{\rm I} \\
                  & = \sum_{\rm I} c_{\rm I} \,\, \left( 1 + \sum_{\mu } t_{\rm I \mu}^{(1)}\,\, T_{\rm I \mu}  \right) \ket{\rm I}
 \end{aligned}
\end{equation}
By introducing the excitation operator $T_{\rm I}^{(1)}$ acting only on $\ket{\rm I}$ as:
\begin{equation}
 \label{def_ti}
 T_{\rm I}^{(1)} = \sum_{\mu} t_{\rm I \mu}^{(1)} \,\, T_{\rm I \mu}
\end{equation}
the expression of $\ket{\Psi^{(1)}}$ in Eq.~\eqref{defPsi1} becomes:
\begin{equation}
 \ket{\Psi^{(1)}} = \sum_{\rm I} c_{\rm I} \left( 1 +  T_{\rm I}^{(1)} \right) \ket{\rm I}
\end{equation}
Such a parameterization for the first-order corrected wave function $\ket{\Psi^{(1)}} $ recalls immediately a first-order Taylor expansion of the 
general JM-MRCC ansatz:
\begin{equation}
 \ket{\rm JM-MRCC} = \sum_{\rm I } c_{\rm I} \,\, e^{T_{\rm I}} \,\, \ket{\rm I}
\end{equation}
Also, within the present formalism, the class of the double excitations can be factorized
as shown in the previous section.
Therefore, using the reference-independent amplitudes defined in Eq.~\eqref{eq:tmnpq}  one can define a unique double excitation operator $\mathcal{T}^{(1)}_{D}$ as:
\begin{equation}
 \mathcal{T}^{(1)}_{D} = \sum_{m,n,p,q} {\mathcal{T}_{mn}^{qp}}^{(1)} 
\end{equation}
recalling thus the formalism of the internally-contracted-MRCC\cite{icmrcc_evangelista,icmrcc_kohn_1,icmrcc_kohn_2,icmrcc_kohn_3} (ic-MRCC) which uses a unique excitation operator $\mathcal{T}$ 
as in the single-reference coupled-cluster: 
\begin{equation} 
 \label{eq_ic_mrcc}
 \begin{aligned}
 \ket{\rm ic-MRCC} & = e^{\mathcal{T}}\ket{\psi^{(0)}} \\
                   & = e^{\mathcal{T}} \sum_{\rm I} c_{\rm I} \,\, \ket{\rm I}
 \end{aligned}
\end{equation} 
However, unlike the ic-MRCC formalism, the JM-MRPT2 equations do not suffer from the linear dependency problems.\cite{icmrcc_kohn_1} 
In such a perspective, as the energy provided by the JM-MRPT2 equations is size-extensive, it can be seen as a linearized coupled cluster version using 
a hybrid parameterization of the wave function: internally contracted ansatz for the double excitation operators and JM ansatz for the single excitation 
operators.

\subsection{Determinant-based multi-reference perturbation theories}
The JM-MRPT2 presented here can be directly compared to the CIPSI method, just as the JM-HeffPT2 can be directly compared to the Shifted-$B_k$ method. 
Indeed, by using the following amplitudes: 
\begin{equation}
 t_{\rm I \mu}^{\rm CIPSI} = \frac{\elemm{\rm I}{H}{\mu}}{e^{(0)} - \elemm{\mu}{H}{\mu}}
\end{equation}
in the equation of the second-order correction on the energy (see Eq.~\eqref{e2pert}), one obtains the CIPSI energy, and by 
introducing $t_{\rm I \mu}^{\rm CIPSI}$  in the 
definition of the dressed Hamiltonian $\tilde{H}^{(2)}$, one obtains the Shifted-$B_k$ energy. 
As mentioned previously, it has been shown that the size-consistency error of these methods comes from the unbalanced treatment between 
the variational energy of a multi-reference wave function such as $\ket{\psi^{(0)}}$ and the variational energy of the single Slater 
determinant $\ket{\mu}$. Such an error is not present within the definitions of the excitation energies in the JM-MRPT2 method as 
the latter introduces expectation values of the Hamiltonian over linear combinations of perturber Slater determinants.  

In a similar context, one can compare the JM-HeffPT2 method to the Split-GAS\cite{splitgas} of Li Manni \textit{et al} 
whose definition of the amplitude is:
\begin{equation}
 t_{\rm I \mu}^{\rm Split-GAS} = \frac{\elemm{\rm I}{H}{\mu}}{e^{(0)} + e^{(2)}  - \elemm{\mu}{H}{\mu}}
\end{equation}
In the Split-GAS framework, the correlation energy $e^{(2)}$ brought by the perturbers is included in the energy denominator, 
which introduces self consistent equations as in the Brillouin-Wigner perturbation theory.\cite{bw} 
However, the size-consistency error in such a method is even more severe than in the Shifted-$B_k$ as the excitation energies are much larger 
due to the presence of the total correlation energy $e^{(2)}$.

\section{Computational cost}
\subsection{Mathematical complexity and memory requirements}
Compared to other size-extensive MRPT2 methods, a clear advantage of the JM-MRPT2 is its simplicity.
The NEVPT2 approach requires to handle the four-body density matrix and the CASPT2 needs to
handle the three-body density matrix. Both of these computationally intensive phases can be skipped 
in our formalism as one only needs to compute expectation values whose number are relatively small compared to NEVPT2 and CASPT2.  
The most involved quantity to be computed is
\begin{equation}
 \Delta E^{(0)}_{ir} = e^{(0)} - \frac{
 \sum_{\rm I} \sum_{\rm J} c_{\rm I} \elemm{\rm I}{H\,\,a^{\dagger}_{r} a_i}{\rm I}\elemm{\rm I}{H}{\rm J} 
 \elemm{\rm J}{H\,\,a^{\dagger}_{r} a_i}{\rm J}   c_{\rm J} } { \left(\sum_{\rm I} c_{\rm I} \elemm{\rm I}{H\,\,a^{\dagger}_{r} a_i}{\rm I}\right)^2}
\end{equation}
for all pairs $(i,r)$ where $i$ is an inactive orbital and $r$ is a virtual orbital. These
quantities need to be only computed once since they can all fit in memory.
Each $\Delta E^{(0)}_{ir}$ is, from the computational point of view, equivalent to an expectation value over the CASSCF wave function. As all the $\Delta E^{(0)}_{ir}$ are independent, the
computation of these quantities can be trivially parallelized.
Regarding the memory footprint of the JM-MRPT2 method, it scales as ${\cal O}(n_a^3)$ ($n_a$ being the
number of active orbitals) for the storage of the 
$\Delta E^{(0)}_{a^{\dagger}_a a^{\dagger}_b a_c}$ and $\Delta E^{(0)}_{a^{\dagger}_a a_b a_c}$ quantities.

Regarding the complexity of the equations for the amplitudes, it is clear that once computed the active part of the denominator, 
the JM-MRPT2 is just a simple sum of contributions. 
This is in contrast with the UGA-SSMRPT2 equations which involve the handling of coupled amplitude equations. 

\subsection{Removal of the determinant-based computational cost}
The present formalisms are formally determinant-based methods, which implies that the computational cost should be proportional to 
the number of perturbers $\ket{\mu}$ that one has to generate to compute the corrections to the energy or the dressing of the Hamiltonian matrix, 
just as in the CIPSI, Shifted-$B_k$ or UGA-SSMRPT2 methods. 
To understand the main computational costs, one can divide the excitation classes according to the difference dedicated CI (DDCI) framework,\cite{ddci}
which classifies the Slater determinants in terms of numbers of holes in the doubly occupied orbitals and particles in the virtual orbitals. 
If $N_{\rm CAS}$ is the number of Slater determinants of the CAS-CI zeroth order wave function, $n_o$, $n_a$ and $n_v$ respectively the number of doubly occupied, 
active and virtual orbitals, one can then classify each excitation class according to the number of perturbers needed to compute their 
contribution to the second-order perturbation correction to the energy:
\begin{enumerate}
 \item the \emph{two-holes-two-particles} excitation class (2h2p) which scales as $N_{\rm CAS} \times n_o^2 \times n_v^2$ 
 \item the \emph{one-hole-two-particles} excitation class (1h2p) which scales as $N_{\rm CAS} \times n_o \times n_a \times n_v^2$ 
 \item the \emph{two-holes-one-particle} excitation class (2h1p) which scales as $N_{\rm CAS} \times n_o^2 \times n_a \times n_v$ 
 \item the \emph{two-particles} excitation class (2p) which scales as $N_{\rm CAS} \times n_v^2 $ 
 \item the \emph{two-holes} excitation class (2h) which scales as $N_{\rm CAS} \times n_o^2 $ 
 \item the \emph{one-hole-one-particle} excitation class (1h1p) which scales as $N_{\rm CAS} \times n_o\times n_v $ 
 \item the \emph{one-particle} excitation class (1p) which scales as $N_{\rm CAS} \times n_v $ 
 \item the \emph{one-hole} excitation class (1h) which scales as $N_{\rm CAS} \times n_o $ 
\end{enumerate}
Nevertheless, our formalism presents several mathematical simplifications that allow one to basically remove any browsing over the Slater determinants 
$\ket{\mu}$, and once more there is a difference between the single and double excitations processes. 

\subsection{Factorization of the most numerous double excitation processes}
As the five most computationally demanding excitation classes involve only double excitation operators in their equations, 
their contribution can be formalized directly thanks to the one- and two-body density matrices of the zeroth-order wave function. 
To understand how, one can write the second-order correction to the energy as:
\begin{equation}
 \begin{aligned}
 & e^{(2)}_{\text{double exc.}} =\\& \sum_{m,\,n,\,p,\,q}  \sum_{\rm I} c_{\rm I} \bra{\psi^{(0)}} H \,\, a^{\dagger}_q a^{\dagger}_p a_n a_m\ket{\rm I} \frac{((mq|np))}{\Delta E^{(0)}_{a^{\dagger}_q a^{\dagger}_p a_n a_m}} \\
                              \\& = \sum_{m,\,n,\,p,\,q}  \sum_{\rm I,\,\, J} c_{\rm I} c_{\rm J} \bra{\rm J} H \,\, a^{\dagger}_q a^{\dagger}_p a_n a_m\ket{\rm I} \frac{((mq|np))}{\Delta E^{(0)}_{a^{\dagger}_q a^{\dagger}_p a_n a_m}}
 \end{aligned}
\end{equation}
Consequently, as $\bra{\rm J} H \,\, a^{\dagger}_q a^{\dagger}_p a_n a_m\ket{\rm I}$ is necessarily of type:
\begin{equation}
 \bra{\rm J} H \,\, a^{\dagger}_q a^{\dagger}_p a_n a_m\ket{\rm I} = ((ef|gh)) \bra{\rm J}  a^{\dagger}_f a^{\dagger}_h a_g a_e  \,\, a^{\dagger}_q a^{\dagger}_p a_n a_m\ket{\rm I}
\end{equation}
one can reformulate the second-order correction to the energy in terms of many-body density matrices:
\begin{equation}
 \begin{aligned}
  e^{(2)}_{\text{double exc.}}  = &\sum_{m,\,n,\,p,\,q, \,e, \, f, \,g,\,h} \bra{\psi^{(0)}}  a^{\dagger}_f a^{\dagger}_h a_g a_e  \,\, a^{\dagger}_q a^{\dagger}_p a_n a_m\ket{\psi^{(0)}} \\ 
                                  &\frac{((mq|np))\,\, ((ef|gh))}{\Delta E^{(0)}_{a^{\dagger}_q a^{\dagger}_p a_n a_m}}
 \end{aligned}
\end{equation}
Such a formulation avoids completely to run over Slater determinants, and consequently kills the prefactor in $N_{\rm CAS}$ involved in 
each of the excitation classes, just as in the internally-contracted formalisms. Of course, because of the restrictions in terms of holes and particles 
in the inactive and virtual orbitals, the handling of the four-body density matrix never occurs in our formalism. 
We report here the explicit equations for the energetic corrections of the five most numerous double excitation classes: 
\begin{equation}
% e^{(2)}_{\rm 2h2p} = \frac{1}{2} \sum_{i,\,j} \frac{3 (iv|jr)^2 + (ir|jv)^2 - 2 (iv|jr)(ir|jv)}{\epsilon_i + \epsilon_j - \epsilon_v - \epsilon_r}
 e^{(2)}_{\rm 2h2p} = \frac{1}{2} \sum_{i,j} \frac{3 (iv|jr)^2 + (ir|jv)^2 - 2 (iv|jr)(ir|jv)}{\epsilon_i + \epsilon_j - \epsilon_v - \epsilon_r}
\end{equation}
\begin{equation}
% e^{(2)}_{\rm 1h2p} = \frac{1}{2} \sum_{i, \,v, \,r} \sum_{a,\, b}\elemm{\psi^{(0)}}{a_a a^{\dagger}_b}{\psi^{(0)}} \frac{((ir|av))((ir|bv))}{\epsilon_i + \Delta E^{(0)}_{a^{\dagger}_a} -\epsilon_r - \epsilon_v }
 e^{(2)}_{\rm 1h2p} = \frac{1}{2} \sum_{i,v,r,a,b}\elemm{\psi^{(0)}}{a_a a^{\dagger}_b}{\psi^{(0)}} \frac{((ir|av))((ir|bv))}{\epsilon_i + \Delta E^{(0)}_{a^{\dagger}_a} -\epsilon_r - \epsilon_v }
\end{equation}
\begin{equation}
% e^{(2)}_{\rm 2h1p} = \frac{1}{2} \sum_{i, \,j, \,r} \sum_{a,\, b}\elemm{\psi^{(0)}}{a^{\dagger}_a a_b }{\psi^{(0)}} \frac{((ir|aj))((ir|bj))}{\epsilon_i + \epsilon_j + \Delta E^{(0)}_{a_a} -\epsilon_r }
 e^{(2)}_{\rm 2h1p} = \frac{1}{2} \sum_{i,j,r,a,b}\elemm{\psi^{(0)}}{a^{\dagger}_a a_b }{\psi^{(0)}} \frac{((ir|aj))((ir|bj))}{\epsilon_i + \epsilon_j + \Delta E^{(0)}_{a_a} -\epsilon_r }
\end{equation}
\begin{equation}
% e^{(2)}_{\rm 2p} = \frac{1}{2} \sum_{r, \,v} \sum_{a,\, b, \, c,\, d}\elemm{\psi^{(0)}}{a^{\dagger}_a a^{\dagger}_b a_c a_d}{\psi^{(0)}} \frac{((ar|bv))((cr|dv))}{\Delta E^{(0)}_{a_c a_d} -\epsilon_r -\epsilon_v }
 e^{(2)}_{\rm 2p} = \frac{1}{2} \sum_{r,v,a,b,c,d}\elemm{\psi^{(0)}}{a^{\dagger}_a a^{\dagger}_b a_c a_d}{\psi^{(0)}} \frac{((ar|bv))((cr|dv))}{\Delta E^{(0)}_{a_c a_d} -\epsilon_r -\epsilon_v }
\end{equation}
\begin{equation}
% e^{(2)}_{\rm 2h} = \frac{1}{2} \sum_{i, \,j} \sum_{a,\, b, \, c,\, d}\elemm{\psi^{(0)}}{a_a a_b a^{\dagger}_c a^{\dagger}_d}{\psi^{(0)}} \frac{((ai|bj))((ci|dj))}{ \epsilon_i + \epsilon_j + \Delta E^{(0)}_{a^{\dagger}_c a^{\dagger}_d}  }
 e^{(2)}_{\rm 2h} = \frac{1}{2} \sum_{i,j,a,b,c,d}\elemm{\psi^{(0)}}{a_a a_b a^{\dagger}_c a^{\dagger}_d}{\psi^{(0)}} \frac{((ai|bj))((ci|dj))}{ \epsilon_i + \epsilon_j + \Delta E^{(0)}_{a^{\dagger}_c a^{\dagger}_d}  }
\end{equation}

\subsection{Simplification for the 1h1p excitation class}
Thanks to the factorization of the most numerous double excitations processes, the remaining main computational cost comes from 
the single excitations involved in the 1h1p excitation class. 
In the case of single excitation processes, the factorization cannot be applied as the Hamiltonian matrix elements depend on the Slater determinant
on which the single excitation is applied. 
The total energetic correction brought by the single excitation processes involved in the 1h1p excitation class can be expressed as follows:
\begin{equation}
 \label{1h1p}
 \begin{aligned}
e^{(2)\,\,\text{Single exc.}}_{\rm 1h1p} & = \sum_{i,\,r} \sum_{\rm I} \elemm{\psi^{(0)}}{H\,\,a^{\dagger}_r a_i}{\rm I} c_{\rm I}\frac{\elemm{\rm I}{H\,\,a^{\dagger}_r a_i}{\rm I}}{\Delta E^{(0)}_{ir}}\\
                                         & = \sum_{i,\,r} \sum_{\rm I,\,\,J} c_{\rm J} \elemm{\rm J}{H\,\,a^{\dagger}_r a_i}{\rm I} c_{\rm I}\frac{\elemm{\rm I}{H\,\,a^{\dagger}_r a_i}{\rm I}}{\Delta E^{(0)}_{ir}}
 \end{aligned}
\end{equation}
As the Hamiltonian matrix elements $\elemm{\rm J}{H\,\,a^{\dagger}_r a_i}{\rm I}$ are simply:
\begin{equation}
 \elemm{\rm J}{H\,\,a^{\dagger}_r a_i}{\rm I} = ((ir|ab)) \elemm{\rm J}{a^{\dagger}_b a_a}{\rm I}
\end{equation}
one can reformulate the sum as:
\begin{equation}
 \begin{aligned}
 e^{(2)\,\,\text{Single exc.}}_{\rm 1h1p} & = \sum_{\rm I,\,\,J} c_{\rm J} c_{\rm I} \,\, \sum_{a,\,b} \mathcal{F}_{ab}^{\rm I}\,\,\elemm{\rm J}{a^{\dagger}_b a_a}{\rm I}
 \end{aligned}
\end{equation}
where the quantity $\mathcal{F}_{ab}^{\rm I}$ is the effective Fock operator associated with the Slater determinant $\ket{\rm I}$ 
involving the active orbitals $a$ and $b$:
\begin{equation}
  \mathcal{F}_{ab}^{\rm I} = \sum_{i,\,r} ((ir|ab)) \frac{\elemm{\rm I}{H\,\,a^{\dagger}_r a_i}{\rm I}}{\Delta E^{(0)}_{ir}}
\end{equation}
Of course, as $\elemm{\rm I}{H\,\,a^{\dagger}_r a_i}{\rm I}$ depends on the occupation of $\ket{\rm I}$, there is one effective Fock operator 
for each reference determinant $\ket{\rm I}$ which would suggest to compute explicitly these quantities for each Slater determinant within the CAS-CI space. 
Nevertheless, one can notice that $\elemm{\rm I}{H\,\,a^{\dagger}_r a_i}{\rm I}$ is just a sum of terms:
\begin{equation}
  \elemm{\rm I}{H\,\,a^{\dagger}_r a_i}{\rm I} = \sum_{m \text{ occupied in }\ket{\rm I}} ((ir|mm)).
\end{equation}
Considering that the inactive orbitals are always doubly occupied in $\ket{\rm I}$, this sum can
be split into an inactive and an active contribution, namely
\begin{equation}
 \elemm{\rm I}{H\,\,a^{\dagger}_r a_i}{\rm I} = F_{ir}^{\text{c.s.}} + F_{ir}^{\rm I} 
\end{equation}
where $F_{ir}^{\text{cs}}$ and  $F_{ir}^{\rm I} $ are defined as
\begin{equation}
 F_{ir}^{\text{cs}} = \sum_{j\,\, \text{doubly occupied in }\ket{\rm I}} 2(ir|jj) - (rj|ij) 
\end{equation}
\begin{equation}
 F_{ir}^{\rm I} = \sum_{c\,\, \text{occupied in} \ket{\rm I}} ((ir|cc))
\end{equation}
Therefore, one can first compute the effective Fock operator associated with the closed shell orbitals:
\begin{equation}
 \mathcal{F}_{ab}^{\text{cs}} = \sum_{i,\,r} ((ir|ab)) \frac{F_{ir}^{\text{cs}}}{\Delta E^{(0)}_{ir}}
\end{equation}
which is common for all the Slater determinants $\ket{\rm I}$ within the CAS-CI space. 
Then, what differentiates the effective Fock operator between two different determinants $\ket{\rm I}$ and $\ket{\rm J}$ is the 
active part:
\begin{equation}
 F_{ab}^{\rm I} =  \sum_{i,\,r} ((ir|ab)) \frac{F_{ir}^{\rm I}}{\Delta E^{(0)}_{ir}}
\end{equation}
One can then notice that the active part of the Fock operator $F_{ir}^{\rm I}$ is just a sum 
over all active orbitals occupied in $\ket{\rm I}$ of quantities that only depend on the active orbitals:
\begin{equation}
 F_{ab}^{\rm I} = \sum_{c\,\, \text{occupied in} \ket{\rm I}}  F_{ab}^c
\end{equation}
where $F_{ab}^c$ is nothing but:
\begin{equation}
 \label{f_abc}
 F_{ab}^c =   \sum_{i,\,r} ((ir|ab)) \frac{((ir|cc))}{\Delta E^{(0)}_{ir}} 
\end{equation}
Therefore, by computing and storing all possible $F_{ab}^c$ together with $\mathcal{F}_{ab}^{\text{cs}}$, 
one can then easily rebuild the total effective operator of a given Slater determinant $\ket{\rm I}$:
\begin{equation}
 \mathcal{F}_{ab}^{\rm I} = \mathcal{F}_{ab}^{\text{cs}} + \sum_{c\,\, \text{occupied in} \ket{\rm I}}  F_{ab}^c
\end{equation}
and consequently compute the total second-order correction to the energy $e^{(2)\,\,\text{Single exc.}}_{\rm 1h1p}$ as a simple expectation value. 
To summarize, a computational step scaling as $N_{\rm CAS}^2 \times n_o \times n_v$ (see Eq.~\eqref{1h1p}) is replaced 
by a first calculation scaling as $n_{act}^3 \times n_o \times n_v$ (see Eq.~\eqref{f_abc}),
followed by the computation of an expectation value scaling as $N_{\rm CAS}^2$,
independent of the number of doubly occupied and virtual orbitals. 

\section{Numerical results}
\label{results}
The present section spells out the numerical results obtained for the potential energy curves and corresponding spectroscopic 
constants of six molecules involving a single, double and triple bond breaking, which are F$_2$, FH, C$_2$H$_6$, C$_2$H$_4$, H$_2$O, and N$_2$.  
A numerical test of strong separability is also provided in the case of the F$_2\dots$FH molecule. 
\subsection{General computational details}
\begin{table}
\caption{Geometries used for the ethane and ethylene molecules.} 
 \begin{center}
 \label{geom}
 \begin{ruledtabular}
 \begin{tabular}{lcc}
 Geometrical parameters &   C$_2$H$_6$   &   C$_2$H$_4$    \\ 
\hline
       C-H (\AA)        &   1.103        &   1.089         \\ 
    H-C-C (\degree)     & 111.2          & 120.0           \\ 
    H-C-H (\degree)     & 107.6          & 120.0           \\ 
   H-C-C-H (\degree)    & 180.0          & 180.0           \\ 
\end{tabular}
\end{ruledtabular}
\end{center}
\end{table}
The cc-pVDZ basis set has been used in all cases, except for the FH molecule for which the aug-cc-pVDZ basis set was retained, 
and pure spherical harmonics were used for all calculations. The frozen core approximation has been used 
and consequently the $1s$ electrons were systematically frozen for all non-hydrogen atoms.  
The near FCI reference values were obtained using the CIPSI algorithm  
developed in the program \textit{Quantum Package}\cite{qp} and all calculations were converged below 0.1 mH. 
The shifted-$B_k$, JM-MRPT2 and JM-HeffPT2 have been implemented in the \textit{Quantum Package}, 
and all CASSCF calculations were performed using the GAMESS(US)\cite{gamess} software.
The CASPT2 calculations were performed with MOLCAS 7.8,\cite{molcas} while the NEVPT2 results were obtained using stand-alone codes developed at the University of Ferrara and interfaced with MOLCAS 7.8.
The geometrical parameters used for the C$_2$H$_6$ and C$_2$H$_4$ molecules can be found in Table \ref{geom}, 
and the H-O-H angle of the H$_2$O molecule has been set to 110.6$\degree$. 

%TODO : ReWRITE
In order to compare the performance of the here proposed formalisms with other determinant-based MPRT2 methods, 
we have also performed calculations using the Shifted-$B_k$ method using an Epstein-Nesbet zeroth order Hamiltonian,  
and we also report results obtained at the Mk-MPRT2\cite{mkmrpt2_4} and UGA-SSMRPT2\cite{ugamrpt2} level of theories when available. 
%TODO : end ReWRITE
For the sake of comparison with other state-of-the-art methods, we also report the spectroscopic constants and the error with respect to FCI 
obtained at the strongly-contracted (SC-NEVPT2) and partially contracted (PC-NEVPT2) NEVPT2 using delocalized orbitals, together with CASPT2 with two different IPEA values. The IPEA values were 
chosen as 0 as in the original formulation of CASPT2, and 0.25 
corresponding to the nowadays standard CASPT2 method.

\subsection{Definition of the active spaces and localized orbitals}
All MRPT2 calculations started with with a minimal valence CASSCF involving the bonding and anti-bonding orbitals 
of each bond being broken along the potential energy curve. 
In the case of the single bond breaking, it simply implies a CASSCF(2,2) with the $\sigma$ and $\sigma^*$ orbitals.  
The following minimal valence active spaces are used for the three systems involving multiple bond breaking:  
for the H$_2$O molecule, a CASSCF(4,4) with  four orbitals 
of valence character (using the $C_{2v}$ symmetry point group, two orbitals of the A$_1$ irrep and two orbitals of the B$_2$ irrep
having a C-H bonding character) ; for the C$_2$H$_4$ molecule, a CAS(4,4) has been performed using 
the bonding and anti-bonding orbitals of both the $\sigma$ and $\pi$ C-C bonds; for the N$_2$ molecule, 
a CAS(6,6) has been used with the bonding and anti-bonding orbitals of the $\sigma$ and the two $\pi$ bonds.  

Nevertheless, as it is the case for many multi-reference perturbation theories, 
our formalism is not invariant through orbital rotations within each orbital space (active, inactive and virtual).
Therefore one can choose to use delocalized orbitals, as the canonical ones, or localized orbitals. 
The present formalism is strictly separable when localized orbitals are used, so it seems therefore natural to use localized active orbitals 
rather than the canonical ones. 
In the case F$_2$, N$_2$, C$_2$H$_6$ and C$_2$H$_4$, these orbitals are simply obtained by a rotation of $\pi/4$ between the bonding and anti-bonding 
active orbitals ($\sigma$ and $\sigma^*$ for the $\sigma$ bond, $\pi$ and $\pi^*$ for the $\pi$ bonds and so on). 
In the case of the FH and H$_2$O molecules, the active orbitals were obtained thanks to a rotation of the canonical active MOs in order to maximize 
the overlap with reference localized orbitals following chemical intuition: for the FH molecule they consist in the $2p_z$ atomic orbital of the fluorine 
and $1s$ atomic orbital of the hydrogen atom, and for the H$_2$O molecule they consist in the two $1s$ atomic orbitals of the hydrogen atoms and 
of two simple linear combinations of the $2p_x$ and $2p_y$ orbitals, each one pointing to a given hydrogen atom. 

Even if the present formalism is strictly separable only using localized orbitals,  
we nevertheless investigate the dependency of the choice of the active orbitals for the three molecules involving a single bond breaking 
(F$_2$, FH and C$_2$H$_6$) for which we report calculations both with canonical delocalized active orbitals (which are referred as ``deloc'') 
and localized active orbitals. 

\subsection{Single bond breaking}

\begin{figure}[h]
  \includegraphics[angle=270,width=0.5\textwidth]{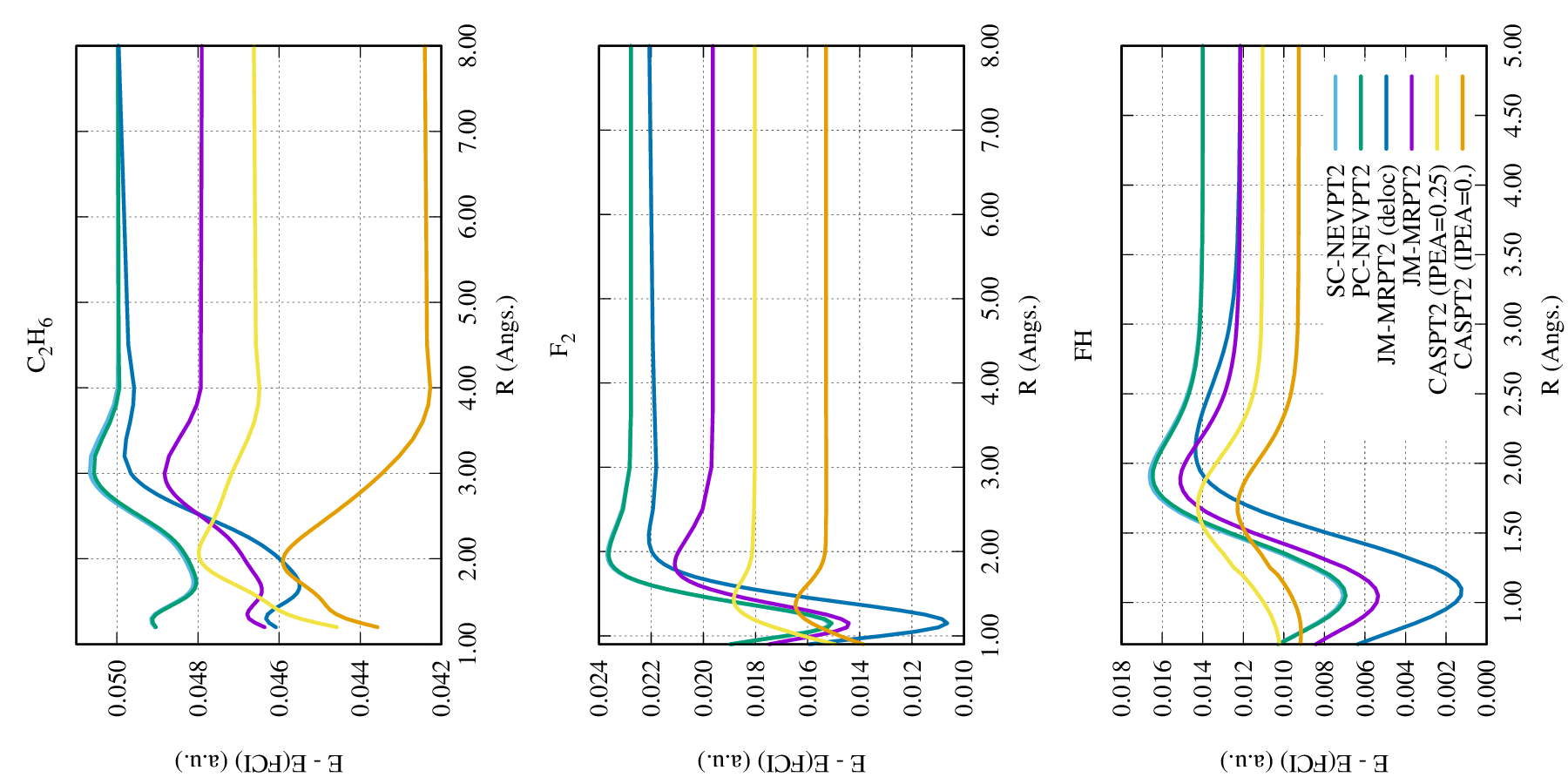}
   \caption{Comparison of different MR-PT2 schemes with the FCI energy along the potential energy curves of
C$_2$H$_6$, F$_2$, with the cc-pVDZ basis set, and FH with the aug-cc-pVDZ basis set. Energy differences in atomic units.}
  \label{fig:pt2_single}
\end{figure}

\begin{table*}
\caption{Non-parallelism errors and spectroscopic constants computed from the potential energy curves obtained at different computational levels for the F$_2$, C$_2$H$_6$ and FH molecules. 
NPE and $D_0$ are reported in mH,  $R_{\rm eq}$ in \AA and $k$ in Hartree/\AA$^2$.}
 \label{second_order}
 \begin{center}
 \begin{ruledtabular}
 \begin{tabular}{lcccccccccccc}
%\hline
                        &\multicolumn{4}{c}{F$_2$}          &\multicolumn{4}{c}{C$_2$H$_6$}     &\multicolumn{4}{c}{FH}              \\
                        & NPE  & $D_0$ & $R_{\rm eq}$ & $k$  & NPE  & $D_0$ & $R_{\rm eq}$ & $k$  & NPE  & $D_0$ & $R_{\rm eq}$ & $k$   \\ 
\hline
         CASSCF         & 30.7 &  22.1 &     1.53     & 0.43 & 27.7 & 154.0 &     1.55     & 0.99 & 35.3 & 180.0 &     0.92     & 2.15  \\ 
\hline
        JM-MRPT2        & 6.7  &  46.3 &     1.44     & 0.85 & 2.5  & 179.0 &     1.53     & 1.06 & 9.7  & 220.4 &     0.93     & 2.11  \\ 
   JM-MRPT2  (deloc)    & 11.4 &  51.1 &     1.43     & 0.93 & 4.5  & 181.6 &     1.54     & 1.06 & 13.1 & 224.3 &     0.93     & 2.13  \\ 
       SC-NEVPT2        & 8.5  &  48.1 &     1.44     & 0.88 & 2.6  & 179.2 &     1.54     & 1.07 & 9.5  & 220.5 &     0.93     & 2.11  \\ 
       PC-NEVPT2        & 8.5  &  48.2 &     1.44     & 0.88 & 2.5  & 179.2 &     1.54     & 1.07 & 9.5  & 220.5 &     0.93     & 2.11  \\ 
    CASPT2 (IPEA=0.)    & 2.6  &  44.1 &     1.46     & 0.74 & 3.6  & 175.0 &     1.53     & 1.08 & 3.1  & 214.1 &     0.92     & 2.16  \\ 
   CASPT2 (IPEA=0.25)   & 3.9  &  44.3 &     1.46     & 0.75 & 3.4  & 177.8 &     1.53     & 1.09 & 4.0  & 214.5 &     0.92     & 2.17  \\ 
 Mk-MRPT2          $^b$ &  -   &  47.2 &     1.44     & 0.60 &  -   &   -   &      -       &  -   &  -   &   -   &      -       &   -   \\ 
 Mk-MRPT2   (deloc)$^b$ &  -   &  48.4 &     1.44     & 0.71 &  -   &   -   &      -       &  -   &  -   &   -   &      -       &   -   \\ 
\hline
       JM-HeffPT2       & 7.4  &  50.1 &     1.45     & 0.87 & 2.4  & 179.4 &     1.53     & 1.05 & 8.9  & 221.9 &     0.93     & 2.11  \\ 
   JM-HeffPT2 (deloc)   & 14.2 &  56.2 &     1.44     & 0.94 & 5.4  & 182.2 &     1.54     & 1.05 & 14.5 & 226.1 &     0.94     & 2.14  \\ 
     Shifted $B_k$      & 6.6  &  50.1 &     1.48     & 0.80 & 8.6  & 136.4 &     1.64     & 0.75 & 44.3 & 216.4 &     0.93     & 2.11  \\ 
 Shifted $B_k$ (deloc)  & 5.7  &  91.8 &     1.41     & 1.26 & 4.7  & 220.5 &     1.53     & 1.09 & 26.7 & 236.1 &     0.94     & 2.31  \\ 
\hline
        FCI$^a$         &  -   &  45.1 &     1.46     & 0.77 &  -   & 177.7 &     1.53     & 1.06 &  -   & 214.4 &     0.92     & 2.16  \\ 
  \end{tabular}
  \end{ruledtabular}
 \addtolength{\tabcolsep}{-8pt}
 \end{center}
\begin{itemize}
\item [$^a$] Results obtained with CIPSI calculations converged up to a second-order perturbative correction lower than $10^{-4}$ Hartree. 
\item [$^b$] Results from Ref \cite{mkmrpt2_4} 
\end{itemize}
 \end{table*}

Table \ref{second_order} presents the spectroscopic constants, namely equilibrium distance ($R_{\rm eq}$), the bond energy ($D_0$) and the second derivative ($k$)
at $R_{\rm eq}$, for the F$_2$, C$_2$H$_6$ and FH molecules at different computational levels. Also, we represent in Figure~\ref{fig:pt2_single}
the difference to the FCI energy along the potential energy curves of those systems.
From these data, several trends can be observed, both regarding the quality of the potential
energy curves 
and the dependency on the choice of the active orbitals. 

\subsubsection{Dependency on the locality of the active orbitals}
From the error of the potential energy curve to the FCI reference,
it appears that the JM-MRPT2 method gives systematically better spectroscopic constants and
a lower 
error with respect to the Full-CI energy when localized orbitals are chosen.
This is consistent with the fact that these methods are strictly separable when localized
orbitals are used.
Therefore, from now on we shall only refer to the results obtained with localized orbitals. 
One can remark that in the case of the F$_2$ molecule where Mk-MRPT2 calculations are available
in the literature,\cite{mkmrpt2_4} the JM-MRPT2 method gives very similar results. 

\subsubsection{Quality of the potential energy curves}
From Table \ref{second_order} it is striking to observe that the spectroscopic constants obtained using the JM-MRPT2 level of theory 
are extremely close to the FCI results. The largest deviation on $D_0$ is of 6 mH for the FH molecule, representing less than 3$\%$ of 
error on the total binding energy, whereas it is of 1.2 mH and 1.3 mH which represents an error of less than 3$\%$ and 1$\%$ on the 
binding energy for the F$_2$ and C$_2$H$_6$ molecules, respectively. The equilibrium geometries obtained at the JM-MRPT2 level 
are always within 1$\%$ of error with respect to the FCI estimates, and so are the $k$ values except for the F$_2$ molecule 
for which a significant deviation of 10$\%$ is observed.  
Except for the quality of the results, one can observe a systematic overestimation of the binding energy at the JM-MRPT2 level. 

The non parallelism error (NPE) is, within the computed points, the difference between the maximum and minimum absolute 
errors with respect to FCI energies. In addition the to the spectroscopic constants, the NPE is also a good indicator of
the quality of the results of a given method. 
Using localized orbitals, the NPE obtained at the JM-MRPT2 is of 6.7~mH for the F$_2$ molecule, 
2.5~mH for C$_2$H$_6$, and 9.7~mH for the FH molecule. The maximum NPE is then for the FH molecule, which has also the largest 
energetic variation among the three molecules studied here.

\subsection{Numerical results for double and triple bond breaking}

\begin{figure}
  \includegraphics[angle=270,width=0.5\textwidth]{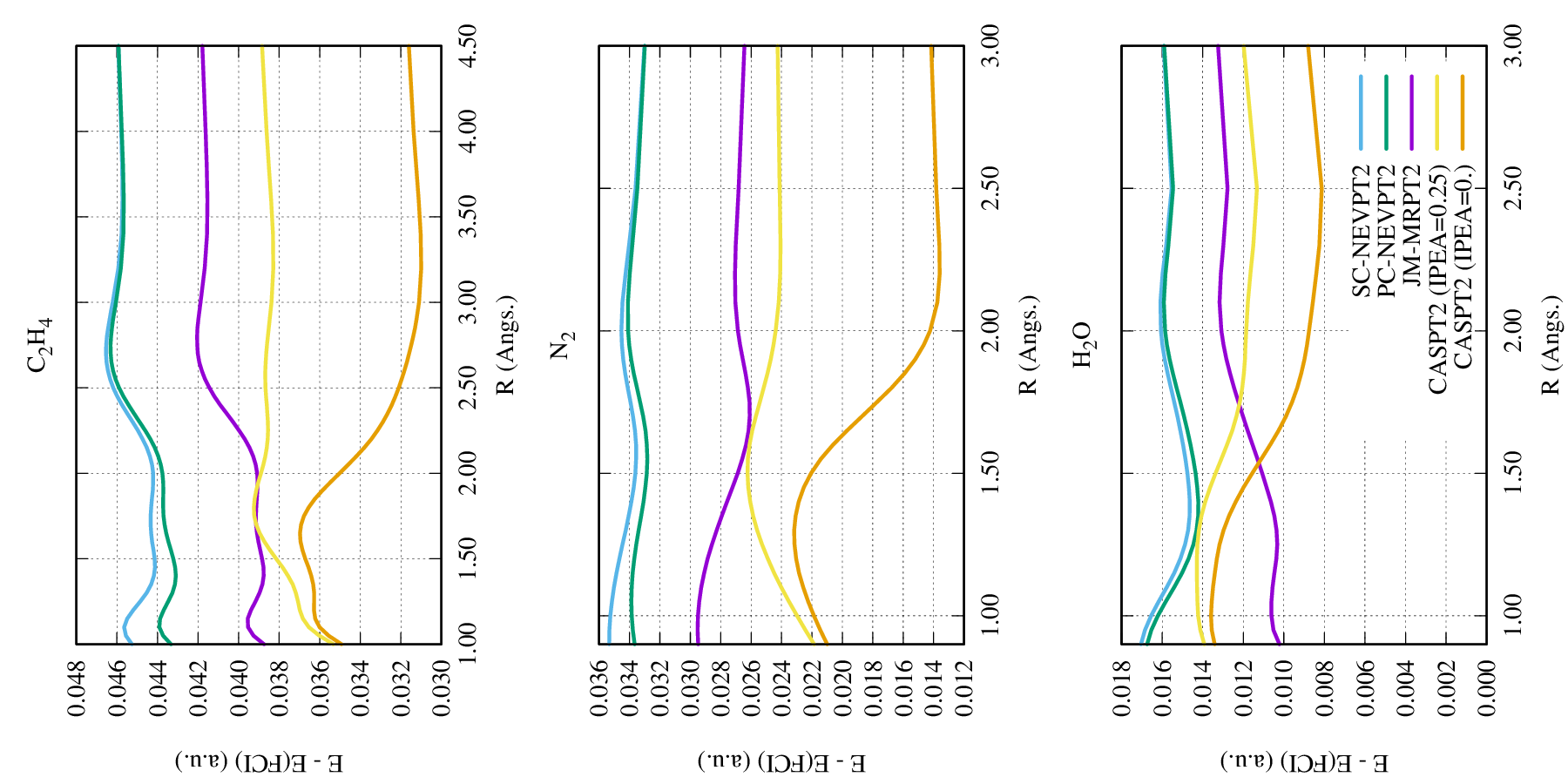}
   \caption{Comparison of different MR-PT2 schemes with the FCI energy along the potential energy curves of
C$_2$H$_4$, N$_2$ and H$_2$O with the cc-pVDZ basis set. Energy differences in atomic units.}
  \label{fig:pt2_multiple}
\end{figure}

\begin{table*}
\caption{Non-parallelism errors and spectroscopic constants computed from the potential energy curves obtained at different computational levels for the H$_2$O, C$_2$H$_4$ and N$_2$ molecules. 
NPE and $D_0$ are reported in mH,  $R_{\rm eq}$ in \AA and $k$ in Hartree/\AA$^2$.} 
 \begin{center}
 \label{double_triple}
 \begin{ruledtabular}
 \begin{tabular}{lcccccccccccc}
                    &\multicolumn{4}{c}{H$_2$O}         &\multicolumn{4}{c}{C$_2$H$_4$}      &\multicolumn{4}{c}{N$_2$}    \\
                    & NPE  & $D_0$ & $R_{\rm eq}$ & $k$  & NPE  & $D_0$ & $R_{\rm eq}$ & $k$  & NPE  & $D_0$ & $R_{\rm eq}$ & $k$   \\ 
\hline
       CASSCF       & 40.9 & 289.3 &     0.96     & 3.74 & 26.2 & 252.6 &     1.36     & 2.03 & 18.2 & 313.7 &     1.11     & 5.34  \\ 
\hline
      JM-MRPT2      & 3.0  & 332.7 &     0.96     & 3.89 & 3.7  & 279.5 &     1.35     & 2.07 & 3.4  & 316.9 &     1.12     & 5.05  \\ 
     SC-NEVPT2      & 2.4  & 329.2 &     0.96     & 3.81 & 2.4  & 278.2 &     1.36     & 2.09 & 2.3  & 317.2 &     1.12     & 5.10  \\ 
     PC-NEVPT2      & 2.5  & 329.5 &     0.96     & 3.81 & 3.2  & 279.3 &     1.35     & 2.10 & 1.3  & 318.2 &     1.12     & 5.10  \\ 
  CASPT2 (IPEA=0.)  & 5.5  & 325.4 &     0.96     & 3.86 & 6.0  & 271.9 &     1.35     & 2.10 & 9.6  & 310.2 &     1.12     & 5.07  \\ 
 CASPT2 (IPEA=0.25) & 3.0  & 327.9 &     0.96     & 3.86 & 4.5  & 278.0 &     1.35     & 2.11 & 4.4  & 318.8 &     1.12     & 5.14  \\ 
\hline
     JM-HeffPT2     & 4.8  & 333.9 &     0.96     & 3.85 & 4.0  & 280.2 &     1.35     & 2.08 & 4.5  & 317.1 &     1.12     & 4.99  \\ 
   Shifted $B_k$    & 30.8 & 304.3 &     0.98     & 3.37 & 7.6  & 238.5 &     1.40     & 1.73 & 5.9  & 277.7 &     1.14     & 4.42  \\ 
\hline
      FCI$^a$       &  -   & 330.3 &     0.96     & 3.89 &  -   & 277.0 &     1.35     & 2.09 &  -   & 319.4 &     1.12     & 5.04  \\ 
  \end{tabular}
  \end{ruledtabular}
 \end{center}
\begin{itemize}
\item [$^a$] Results obtained with CIPSI calculations converged up to a second-order perturbative correction lower than $10^{-4}$ Hartree. 
\end{itemize}
 \end{table*}

Table \ref{double_triple} presents the spectroscopic constants obtained for the H$_2$O, C$_2$H$_4$ and N$_2$ molecules and 
Figure~\ref{fig:pt2_multiple} shows the difference to the FCI energy along the potential energy curves.
From Table \ref{double_triple}, it appears that the results obtained with the JM-MRPT2 method follows a trend similar 
to what has been observed with the study of the three molecules involving a single bond breaking: the spectroscopic constants obtained at this levels 
of theory are globally in good agreement with the FCI ones, the $D_0$ obtained at the JM-MRPT2 level tends to be overestimated. 
Also, the absolute error on $D_0$ obtained at the JM-MRPT2 is quite constant: 
2.4~mH, 2.4~mH and 2.5~mH, representing 0.7$\%$, 0.9$\%$ and 0.8$\%$ of the total binding energy for the H$_2$O, C$_2$H$_4$ and N$_2$ molecules, 
respectively.

Regarding the curves displaying errors with respect to the FCI energies, it appears that the
JM-MRPT2 curves are smooth and do not present any intruder state problems, with an NPE between 3 and 4~mH.

\subsection{Comparison of JM-HeffPT2 with Shifted-$B_k$}

\begin{figure*}
  \includegraphics[angle=270,width=\textwidth]{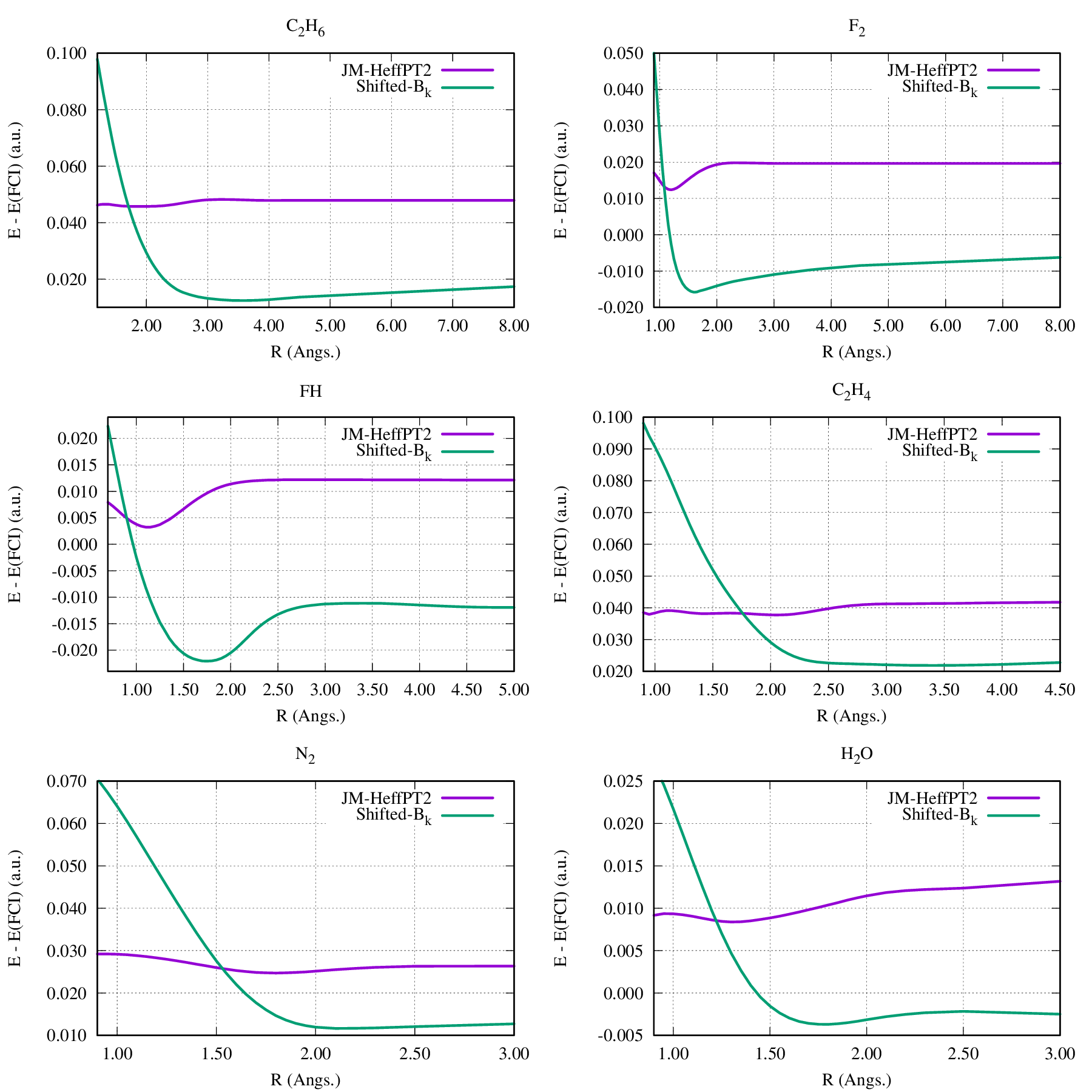}
   \caption{Energy difference of JM-HeffPT2 and Shifted-$B_k$ with respect to the FCI energy along the potential energy curves of
C$_2$H$_6$, F$_2$, C$_2$H$_4$, N$_2$ and H$_2$O with the cc-pVDZ basis set,
and FH with the aug-cc-pVDZ basis set. Energy differences in atomic units.}
  \label{fig:heff}
\end{figure*}

Figure~\ref{fig:heff} shows the difference to the FCI for all the previously
studied systems, for the JM-HeffPT2 and the Shifted-$B_k$ methods. It is clear
that in all the cases the potential energy curves obtained with the JM-HeffPT2
are much more parallel to the FCI curve than the Shifted-$B_k$ ones.
Also, it is worth mentioning that the
JM-HeffPT2 curves are smooth and do not present any intruder state problems.
The spectroscopic constants and NPEs calculated with both methods are given in
tables~\ref{second_order} and~\ref{double_triple}.

\begin{table}
\caption{Ratios $c_i/c_n$ at different internuclear distances, for the F$_2$ molecule (cc-pVDZ).}
\label{tab:ratio}
\begin{ruledtabular}
\begin{tabular}{lccc}
   F2          &      1.4119 \AA  &       2 \AA     &       3 \AA      \\ 
\hline
     CAS-CI    & 0.572          & 0.212          & 0.024           \\ 
   JM-HeffPT2  & 0.646          & 0.273          & 0.033           \\ 
 Shifted-$B_k$ & 0.707          & 0.274          & 0.030           \\ 
\hline
    CIPSI        &      0.638      &      0.259     &      0.030      \\ 
    $N_{\rm det}$ &   \textit{6 321 822}      &    \textit{7 889 806}     &    \textit{12 748 141}     \\ 
\end{tabular}
\end{ruledtabular}
\end{table}

\begin{table}
\caption{Dipole moment (reported in a.u.$^2$) along the internuclear axis obtained at various computational 
levels for the FH molecule (aug-cc-pVDZ).}
\label{tab:dipole}
\begin{ruledtabular}
\begin{tabular}{lccc}
   FH               &      0.95   \AA  &      1.4 \AA     &       1.9 \AA      \\ 
\hline              
     CAS-CI         &      1.07      &      1.87      &     3.20                \\ 
   JM-HeffPT2       &      1.04      &      1.70      &     2.88                \\ 
 Shifted-$B_k$      &      1.01      &      1.47      &     2.19                \\ 
\hline
CIPSI-proj-CAS      &      1.05       &     1.71       &    2.92                \\ 
    $N_{\rm det}$ &   \textit{2 677 789}     &    \textit{2 545 448}       &    \textit{2 153 580}        \\ 
\end{tabular}
\end{ruledtabular}
\end{table}

In general, the energetic values obtained after diagonalizing the effective Hamiltonian
are not better than those obtained with the JM-MRPT2 method.
But the main advantage of JM-HeffPT2 over JM-MRPT2 is that it provides
improved CI-coefficients on the reference space, like the Shifted-$B_k$ method.
To illustrate the quality of the improved wave functions, we report in
table~\ref{tab:ratio} the ratios $c_i/c_n$ where $c_i$ and $c_n$ are the
CI-coefficients of the determinants relative to the ionic and neutral
structures of F$_2$ obtained at the CAS-CI, JM-HeffPT2, Shifted-$B_k$ and CIPSI levels.
As a reference, CIPSI calculations were carried out in the frozen-core FCI space, and the
number of determinants ($N_{\rm det}$) selected in the variational wave function are given in
table~\ref{tab:ratio}. For such large wave functions, the CI-coefficients on the
reference determinants are expected to be very close to the FCI limit.
Both the JM-HeffPT2 and Shifted-$B_k$ show a significant improvement of the wave function,
and the JM-HeffPT2 is in very good agreement with the FCI especially at the equilibrium
distance. Similarly, we report in table~\ref{tab:dipole} computations of the dipole moment along the internuclear 
axis for the FH molecule, and compare it to values obtained by projecting and normalizing large CIPSI wave functions 
on the CAS-CI space (referred hereafter as CIPSI-proj-CAS). 
Therefore, the dipole moment computed with a given method only depends on the relative coefficients of the four 
Slater determinants belonging to the CAS-CI space. 
From these results, it clearly appears that the JM-HeffPT2 method allows to obtain values for the dipole moment 
that are in excellent agreement with that obtained at the CIPSI-proj-CAS level of theory. 
Also, one can notice a significant improvement of the description of the dipole moment going from the CAS-CI wave function 
to the JM-HeffPT2 wave function, implying that the diagonalization of the dressed Hamiltonian leads to 
coefficients within the CAS-CI space that are closer to the ones of the FCI wave function, 
which is not the case for the Shifted-$B_k$ method.

\subsection{Numerical evidence of strong separability}
The present definitions of the JM-MRPT2 and JM-HeffPT2 respect the property of strong separability when localized orbitals are used. 
A formal proof of the strict separability is given in the appendix.
In order to give a numerical example of the strong separability property, 
we report in Table \ref{testseparable} calculations on F$_2$ (F-F=1.45 \AA), FH (F-H=0.90 \AA), and on the super-system  
of F$_2\dots$FH at an intermolecular distance of 100~\AA. As the two subsystems are different, the orbitals obtained by the CASSCF method are localized 
on each system, which is a necessary condition for the strong separability in our formalism. 
\begin{table*}
\caption{Total energies (a. u.) for the numerical separability check on F$_2\dots$FH. } 
 \label{testseparable}
 \begin{center}
 \begin{ruledtabular}
 \begin{tabular}{lcccc}
                  &        CASSCF        &    Shifted-$B_k$    &  $e^{(2)}$-JM-MRPT2  &      JM-HeffPT2       \\ 
\hline
      F$_2$       &  -198.746157368569   &    -199.122170300   &  -0.337009510134933  &  -199.0853051551694   \\ 
        FH        &  -100.031754985880   &    -100.289784498   &  -0.230422886638017  &  -100.2624246672967   \\ 
    F$_2$ + FH    &  -298.777912354448   &    -299.411954798   &  -0.567432396772949  &  -299.3477298224660   \\ 
 F$_2$ $\dots$ FH &  -298.777912354443   &    -299.396752116   &  -0.567432396773035  &  -299.3477298224616   \\ 
\hline
  Absolute error (a.u.)  & 5.0$\times 10^{-12}$ & 1.5$\times 10^{-2}$ & 8.6$\times 10^{-14}$ & 4.4$\times 10^{-12}$  \\ 
  Relative error  & 1.7$\times 10^{-14}$ & 5.1$\times 10^{-5}$ & 1.5$\times 10^{-13}$ & 1.4$\times 10^{-14}$  \\ 
  \end{tabular}
  \end{ruledtabular}
 \end{center}
 \end{table*}
From Table~\ref{testseparable}, it appears that the deviations on the computed correlated energy $e^{(2)}$-JM-MRPT2 (see Eq.~\eqref{e2pert})  
between the super system with non interacting fragments and the sum of the two systems is lower than $10^{-13}$ Hartree, 
which is actually smaller than the non additivity of the CASSCF energies.
For the JM-HeffPT2, the relative error remains in the same order of magnitude than the for the CASSCF.
This shows that no non-separability error was introduced in the effective Hamiltonian.
Finally, one should notice the strong non-separability error of the shifted-$B_k$ approach.

\section{Conclusions and perspectives.\label{concl}}

\subsection{Summary of the main results}
The present work has presented a new MRPT2 approach, the JM-MRPT2 method, 
that uses individual Slater determinants as perturbers and allows for an intermediate Hamiltonian formulation, which is the JM-HeffPT2 approach. 
These methods are strictly size-consistent when localized orbitals are used, as
has been numerically illustrated here. The link of these two new methods with other existing multi-reference theories has been 
established, specially in the case of the SC-NEVPT2 level of theory.
The accuracy of the methods has been investigated on a series of ground state potential energy 
curves up to the full dissociation limit for a set of six molecules involving single (F$_2$, FH, C$_2$H$_6$), double (H$_2$O, C$_2$H$_4$) 
and triple bond breaking (N$_2$), using the cc-pVDZ basis set and the aug-cc-pVDZ basis set in the case of FH.
The two methods proposed here have been compared to near FCI energies thanks to large CIPSI calculations converged bellow 0.1~mH, 
whose values can be found in the supporting information file. 
The quality of the results has been investigated by means of the non-parallelism error and three spectroscopic constants ($R_{\rm eq}$, $D_0$ and $k$) together with 
absolute errors with respect to FCI energies along the whole potential energy curves. 
Among the six molecules studied here, the largest error found on the binding energy at the JM-MRPT2 level of theory 
is of 6~mH for the FH molecule, representing a deviation lower than 3$\%$ with respect to the FCI value. 
In all other cases, the errors on $D_0$ are much smaller, ranging from 1.3~mH to 2.5~mH, which represents deviations between 1$\%$ and 3$\%$ 
with respect to the FCI estimates. The  equilibrium distance is also found to be always within $1\%$ of the FCI values. 
These results are very encouraging, specially considering the simplicity of this second-order perturbation theory, 
and its low computational cost. 
Regarding the JM-HeffPT2 method, its intermediate Hamiltonian formulation allows to take into account the dominant part of the coupling 
between the static and dynamic correlation effects. From what has been observed in the present calculations, 
the diagonalization of the symmetrized intermediate Hamiltonian yields improved CI-coefficients on the reference determinants, together
with a very small NPE compared to the Shifted-$B_k$ method.

\subsection{Perspectives}
Due to its flexibility, the present formalism offers a broad field of perspectives.
First, the JM-MRPT2 and JM-HeffPT2 methods can be formalized with a zeroth order wave function that does 
not need to be a CAS-CI eigenvector. This opens the way of treating much larger active spaces as one can select the dominant configurations 
of a given CAS-CI space thanks to the use of a perturbative criterion (as in the CIPSI algorithm) or by using localized orbitals. 
Second, the reasons of the systematic slight overestimation of the binding energy at the JM-MRPT2 level of theory can also be investigated, 
taking benefit from localized active orbitals and of the clear reading of the reference wave function that they offer. 
Moreover, this allows one to use as zeroth-order wave function quasi diabatic states obtained, for instance, 
by a unitary transformation of a few CI eigenvectors\cite{ovb} (either of a CAS-CI or from a more general CI)
Also, as it has been shown that the present formulation is connected to multi-reference coupled-cluster formalisms, it is possible
to derive the working equations starting from the JM-MRCC ansatz.
This will allow to obtain higher order terms which may correct the slight overestimation of the binding energies.
The coupling of the present formalism with multi-reference coupled cluster models follows naturally.
For instance, the treatment of the most numerous excitation classes at the JM-MRPT2 level can easily be combined 
with the recently introduced JM-MRCC ansatz of some of the present authors.\cite{lambda_mrcc} 
This will allow for a drastic lowering of the computational costs of the JM-MRCC ansatz, and opens the way to the treatment of larger  
systems at high level of \textit{ab initio} theory.

{\it Acknowledgments.}
This work was performed using HPC resources from CALMIP
(Toulouse) under allocation 2016-0510 and from GENCI
(Grant 2016-081738).

\section*{Appendix: proof of separability}
\label{annex}
The present section proposes an analytical proof of strong separability of the JM-MRPT2 method. 
In a MRPT2 framework, the strong separability requires that an excitation $T_A$ located on a system $A$ gives the same 
contribution to the correlation energy with or without the presence of another system $B$ whose zeroth-order wave function 
contains correlation effects. 
To be more specific, let us define the zeroth-order wave function and energy of a system $A$:
\begin{equation}
 \label{psia}
 \ket{\psi^{(0)\,A}} = \sum_{\rm I_{A}} c_{\rm I_A} \ket{\rm I_A}
\end{equation}
\begin{equation}
 \label{e_a}
 E^{(0)\,A} = \frac{\bra{\psi^{(0)\,A}} H_A \ket{\psi^{(0)\,A}}}{\bra{\psi^{(0)\,A}} \psi^{(0)\,A}\rangle}
\end{equation}
and the same quantities for the system $B$:
\begin{equation}
 \label{psib}
 \ket{\psi^{(0)\,B}} = \sum_{\rm I_{B}} c_{\rm I_B} \ket{\rm I_B}
\end{equation}
\begin{equation}
 \label{e_b}
 E^{(0)\,B} = \frac{\bra{\psi^{(0)\,B}} H_B \ket{\psi^{(0)\,B}}}{\bra{\psi^{(0)\,B}} \psi^{(0)\,B}\rangle}
\end{equation}
Let us consider now a given excitation $T_A$ acting only on a system $A$. According to the definition of Eq.~\eqref{def_deltae_uniq}, 
the corresponding contribution to the first-order perturbed wave function is:
\begin{equation}
 \ket{\psi^{(1)\,A}_{T_A}} =\frac{1}{\Delta E^{(0)\,A}_{T_A} }\,\,\ket{\tilde{\psi}^{(1)\,A}_{T_A}} 
\end{equation}
\begin{equation}
 \ket{\tilde{\psi}^{(1)\,A}_{T_A}} = \sum_{\rm I_A} c_{\rm I_A} \,\,\bra{\rm I_A} H_A \,\,T_A\ket{\rm I_A} \,\, T_A\ket{\rm I_A}
\end{equation}
and the excitation energy $\Delta E^{(0)\,A}_{T_A}$ characteristic of the excitation $T_A$ is defined according to Eq.~\eqref{def_deltae} as:
\begin{equation}
 \Delta E^{(0)\,A}_{T_A} = E^{(0)\,A} - \frac{\elemm{\tilde{\psi}^{(1)\,A}_{T_A}}{H_A}{\tilde{\psi}^{(1)\,A}_{T_A}}}{\bra{\tilde{\psi}^{(1)\,A}_{T_A}} \tilde{\psi}^{(1)\,A}_{T_A} \rangle  }
\end{equation}
Therefore, its contribution to the correlation energy of $A$ is:
\begin{equation}
 \begin{aligned}
   e^{(2)\,A}_{T_A} & = \elemm{\psi^{(0)\,A}}{H}{\psi^{(1)\,A}_{T_A}} \\
                    & = \frac{\elemm{\psi^{(0)\,A}}{H_A}{\tilde{\psi}^{(1)\,A}_{T_A}}}{\Delta E^{(0)\,A}_{T_A} }
 \end{aligned}
\end{equation}
The strong separability property will be respected if the contribution to the correlation energy $e^{(2)\,A+B}_{T_A}$ of $T_A$ 
in the case of the super-system non interacting $A\dots B$ is strictly equal to $e^{(2)\,A}_{T_A}$. 
In such a case, the zeroth-order wave function must be the product of the zeroth-order wave function of the two 
sub-systems $A$ and $B$:
\begin{equation}
 \ket{\psi^{(0)\,A+B}} = \ket{\psi^{(0)\,A}} \otimes \ket{\psi^{(0)\,B}}
\end{equation}
which ensures that its corresponding zeroth-order energy is the sum of zeroth-order energies of the sub-systems $A$ and $B$:
\begin{equation}
 \begin{aligned}
 E^{(0)\,A+B} & = \frac{\elemm{\psi^{(0)\,A+B}}{H_A + H_B}{\psi^{(0)\,A+B}} }{\langle\psi^{(0)\,A+B} \ket{\psi^{(0)\,A+B}}} \\
              &= \frac{ \elemm{\psi^{(0)\,A}}{H_A }{\psi^{(0)\,A}} \langle\psi^{(0)\,B} \ket{\psi^{(0)\,B}}}{\langle\psi^{(0)\,A} \ket{\psi^{(0)\,A}} \langle\psi^{(0)\,B} \ket{\psi^{(0)\,B}} }\\
              &+ \frac{ \elemm{\psi^{(0)\,B}}{H_B }{\psi^{(0)\,B}} \langle\psi^{(0)\,A} \ket{\psi^{(0)\,A}}}{\langle\psi^{(0)\,B} \ket{\psi^{(0)\,B}} \langle\psi^{(0)\,A} \ket{\psi^{(0)\,A}} }\\
              & = E^{(0)\,A}  + E^{(0)\,B}
 \end{aligned}
\end{equation}
as the total Hamiltonian can be written as the sum of $H_A$ acting only on the orbitals of $A$ and the corresponding $H_B$  
acting only on the orbitals of $B$. A CAS-CI wave function respects of course such a property. 

Starting from $\ket{\psi^{(0)\,A+B}}$, one can generate the contribution to the first-order perturbed wave function  
$\ket{\psi^{(1)\,A}_{T_A}}$ associated with $T_A$ in the super-system $A\dots B$:
\begin{equation}
 \ket{\psi^{(1)\,A+B}_{T_A}} =\frac{1}{\Delta E^{(0)\,A+B}_{T_A} }\,\,\ket{\tilde{\psi}^{(1)\,A+B}_{T_A}} 
\end{equation}
\begin{equation}
 \label{defpsi_ab}
 \begin{aligned}
 \ket{\tilde{\psi}^{(1)\,A+B}_{T_A}} = &\sum_{\rm I_A\, I_B} c_{\rm I_A} c_{\rm I_B} \,\,T_A\,\,\ket{\rm I_B} \otimes \ket{\rm I_A} \\
  & \bra{\rm I_A}\otimes \bra{\rm I_B} \left(H_A + H_B\right) \,\,T_A\ket{\rm I_B}\otimes\ket{\rm I_A}
 \end{aligned}
\end{equation}
with the following excitation energy $\Delta E^{(0)\,A+B}_{T_A}$:
\begin{equation}
 \Delta E^{(0)\,A+B}_{T_A} = E^{(0)\,A+B} - \frac{\elemm{\tilde{\psi}^{(1)\,A+B}_{T_A}}{H_A+H_B}{\tilde{\psi}^{(1)\,A+B}_{T_A}}}{\bra{\tilde{\psi}^{(1)\,A+B}_{T_A}} \tilde{\psi}^{(1)\,A+B}_{T_A} \rangle  }
\end{equation}
Then, the contribution of $T_A$ to the correlation energy of the super system $A\dots B$ is simply:
\begin{equation}
 \label{e_corr_ab}
  e^{(2)\,A+B}_{T_A} = \frac{\elemm{\psi^{(0)\,A+B}}{H_A + H_B}{\tilde{\psi}^{(1)\,A+B}_{T_A}} }{\Delta E^{(0)\,A+B}_{T_A}} 
\end{equation}
One can then notice that, as $T_A$ only acts on the orbitals of $A$, one has:
\begin{equation}
 \bra{\rm I_A}\otimes \bra{\rm J_B} \left(H_A + H_B\right) \,\,T_A\ket{\rm J_B}\otimes\ket{\rm I_A} = \bra{\rm J_B} {\rm J_B}\rangle 
 \elemm{\rm I_A}{H_A  \,\,T_A}{\rm I_A} 
\end{equation}
and consequently the zeroth-order can be factorized in the Eq.~\eqref{defpsi_ab}:
\begin{equation}
 \begin{aligned}
 \ket{\tilde{\psi}^{(1)\,A+B}_{T_A}} & = \sum_{\rm I_B} c_{\rm I_B} \ket{\rm I_B} \otimes \sum_{\rm I_A} c_{\rm I_A} \bra{\rm I_A}H_A  \,\,T_A\ket{\rm I_A} \,\,T_A\,\,\ket{\rm I_A} 
  \\
                                     & = \ket{\psi^{(0)\,B}} \otimes \ket{\tilde{\psi}^{(1)\,A}_{T_A}}
 \end{aligned}
\end{equation}
This form for the $\ket{\tilde{\psi}^{(1)\,A+B}_{T_A}} $ is crucial, as it has a product structure, implying that 
it will not suffer from any size consistency and separability issues. 
Indeed, the numerator of Eq.~\eqref{e_corr_ab} simply reduces to:
\begin{equation}
 \elemm{\psi^{(0)\,A+B}}{H_A + H_B}{\tilde{\psi}^{(1)\,A+B}_{T_A}}  = \elemm{\psi^{(0)\,A}}{H_A }{\tilde{\psi}^{(1)\,A}_{T_A}} 
\end{equation}
and the denominator of the same equation~\eqref{e_corr_ab} is then:
\begin{equation}
 \begin{aligned}
 \Delta E^{(0)\,A+B}_{T_A} & = E^{(0)\,A+B} - \frac{\elemm{\tilde{\psi}^{(1)\,A}_{T_A}}{H_A}{\tilde{\psi}^{(1)\,A}_{T_A}}}{\bra{\tilde{\psi}^{(1)\,A}_{T_A}} \tilde{\psi}^{(1)\,A}_{T_A} \rangle  } - E^{(0)\,B} \\
                           & = E^{(0)\,A} - \frac{\elemm{\tilde{\psi}^{(1)\,A}_{T_A}}{H_A}{\tilde{\psi}^{(1)\,A}_{T_A}}}{\bra{\tilde{\psi}^{(1)\,A}_{T_A}} \tilde{\psi}^{(1)\,A}_{T_A} \rangle  }\\
                           & = \Delta E^{(0)\,A}_{T_A} 
 \end{aligned}
\end{equation}
and therefore, 
\begin{equation}
 e^{(2)\,A+B}_{T_A} = e^{(2)\,A}_{T_A} 
\end{equation}
Consequently, the JM-MRPT2 is strictly separable provided that a partition of the Hamiltonian in terms of $H_A$ and $H_B$ can be done, 
which supposes local orbitals.

\end{document}